%% file: main__11_.tex
\documentclass[11pt]{article}

\usepackage[margin= 1in]{geometry}
\usepackage[T1]{fontenc}
\usepackage{lmodern}
\usepackage[nopatch=footnote]{microtype}
\usepackage{graphicx}
\usepackage{booktabs}
\usepackage{longtable}
\usepackage{pdflscape}
\usepackage{array}
\usepackage{adjustbox}
\usepackage{amsmath,amssymb}
\usepackage{natbib}
\bibpunct{[}{]}{,}{a}{,}{,}
\usepackage{hyperref}
\usepackage[nameinlink,noabbrev]{cleveref}
\usepackage{subcaption}
\usepackage{float}
\usepackage[section]{placeins}
\usepackage{xcolor}
\usepackage{tikz}
\usetikzlibrary{positioning,arrows.meta,fit,calc}
\usepackage{enumitem}
\usepackage{setspace}
\usepackage{authblk}

\hypersetup{colorlinks=true,linkcolor=black,citecolor=black,urlcolor=blue}
\graphicspath{{figures/main/panels/}{figures/supplementary/}}
\makeatletter
\@ifundefined{r@fig:s1}{\expandafter\gdef\csname r@fig:s1\endcsname{{S1}{1}}}{}
\makeatother

\title{\textbf{scCTM: Continuous Thought Machine Reconstruction of Glioma Cell Fate Dynamics Attractor Landscapes and Algorithmic Perturbation Signatures}}
\author[1,2,3,*]{Abicumaran Uthamacumaran}

\affil[1]{Department of Pathology and Laboratory Medicine, University of British Columbia, Canada}
\affil[2]{Oxford Immune Algorithmics, Reading, United Kingdom}
\affil[3]{Department of Surgical and Interventional Sciences, McGill University, Montreal, Canada}
\affil[*]{Corresponding author: \texttt{a\_utham@live.concordia.ca}}
\date{}

\begin{document}
\maketitle

\begin{abstract}
Cell fate plasticity dynamics in glioma ecosystems is usually inferred from static single-cell snapshots, leaving a gap between manifold geometry, predictive modelling, and mechanistic intervention ranking. We developed scCTM, a complexity theory and machine learning framework that combines the Sakana AI Continuous Thought Machine (CTM) with autoencoding, a continuous actor-critic policy over synchronization-derived latent states for predicting single-cell fate dynamics. We also incorporated integrated-gradient and decoder-weight explainability, network science, empirical Waddington-like potential landscape (attractor) reconstruction, and post-hoc Block Decomposition Method (BDM) perturbation analysis. The complexity framework was evaluated in two malignant cell secondary analyses: GSE210567 stem-like versus differentiated H3K27M cells (18,490 cells) and longitudinal H3K27M initial versus recurrent disease trajectory (20,000 cells) from Sussman et al. (2026) as a ML binary classification problem. CTM-AE achieved AUROC values of $0.99$ in both. CTM-RL preserved classification almost exactly but did not improve reconstruction, whereas MSE increased in both analyses. Out-of-fold CTM-RL latent states produced class-conditioned empirical potential basins across PCA, UMAP, and PHATE embeddings. Initial-to-recurrence prediction was dominated by \textit{IGFBP7, CLU, LSAMP, APOE, MT-CO3, MT-CO2, CADM2, LHFPL3, MT-CO1, NRXN3}, and \textit{SPARCL1}. The corresponding leading BDM perturbations included \textit{HSP90AA1, CD63, PTGDS, IGFBP7, APOE, DPP6, MMP16, OPCML, FTH1}, and \textit{LHFPL3}. Across the two retained analyses, top-rank integration recurrently highlighted \textit{LSAMP, DPP6, SOX6, QKI, ZBTB20, PTPRZ1, CSMD1, SNTG1}, and \textit{LRP1B}. BDM edge perturbation analysis further identified recurrent high-impact network interactions, including epigenetic/chromatin-associated signatures involving ARID1B and developmental or neuronal-state genes such as AUTS2, CTNND2, DPP6, NRXN1, and CADM2. These perturbation-sensitive edges nominate candidate network interactions that may help regulate or constrain cell-state dynamics across primary and recurrent glioma ecosystems. These in silico findings support CTM synchronization as an interpretable latent-state representation and predictive forecasting of cell fate dynamics, as well as BDM perturbation signatures as a complementary in silico intervention ranking layer for computational and systems medicine.

\end{abstract}

\noindent\textbf{Keywords:} Continuous thought machine; single-cell RNA sequencing; glioma; cell-fate plasticity; Waddington landscape; integrated gradients; algorithmic information dynamics; block decomposition method; reinforcement learning; cancer dynamics; artificial intelligence; systems medicine.

\section{Introduction}

Cell fate is an emergent property of multiscale interacting networks, including transcriptional, epigenetic, signalling, metabolic, bioelectric, and environmental processes \citep{Huang2009}. Waddington's epigenetic landscape provided a conceptual representation of this complex process: developing cells move through a structured multidimensional landscape containing valleys corresponding to comparatively stable phenotypic states (attractors), ridges that constrain state-transitions, and branching regions (bifurcations) at which alternative fates become accessible \citep{Waddington1957}. An attractor is a persistent long-term state or behavioural pattern approached by network dynamics and their emergent cell fate trajectories starting within its basin of attraction \citep{Kauffman1969,Huang2005,Wang2011}. In contemporary dynamical systems terminology, stable or metastable cell states can be treated as attractor-like regions, whereas differentiation, plasticity, malignant transformation, recurrence, and therapeutic reprogramming correspond to movements between their basins of attraction \citep{Huang2009}. This complex systems framework is especially relevant to glioma ecology, in which disrupted developmental programs/differentiation hierarchies, oncogenic lesions, treatment selection pressure, and microenvironmental signals can support multiple coexisting cellular states rather than a single fixed tumour identity \citep{Filbin2018,Neftel2019,Jessa2022,Sussman2026}.

Single-cell or single-nucleus transcriptomics offers high-dimensional profiling of cellular dynamics and heterogeneity but generally measures different cellular identities at static, isolated snapshots rather than following the whole-systems differentiation trajectory of an individual cell or cell fates as state-transitions along an attractor landscape. Trajectory inference methods therefore estimate a pseudotemporal ordering, branching structure, or transition probability from relationships among sampled cellular states. Pseudotemporal ordering, diffusion pseudotime, graph-based approaches, and statistically-bound manifold learning methods have shown that such orderings can recover biologically meaningful developmental continua \citep{Trapnell2014,Haghverdi2016,Moon2019}. Their outputs nevertheless depend on reductionistic assumptions or statistical distributions-based sampling density, root-state specification, simplistic topology, and model assumptions that may fail to capture the nonlinear dynamics, nonequilibrium physics, and collective (emergent) dynamics seen in complex adaptive systems such as glioma ecologies. A geometric path through an embedding, such as morphospace patterning or latent space of AI capable of capturing complex attractor dynamics, is not automatically a biological velocity field, and a high-density region is not automatically an experimentally verified attractor.

Most trajectory inference methods primarily reconstruct geometry from some static snapshots that lack complex dynamics forecasting such as critical transitions or plasticity along a (disrupted) differentiation hierarchy: they position cells often with the assumption that they are stable or fixed point attractors on a manifold and infer likely paths through that manifold. Geometry alone does not identify the generating mechanisms that stabilize a cell state, determine whether a transition is dynamically accessible, or distinguish a central regulatory intervention from a correlated marker. In tumor ecologies, such as high-grade gliomas, there is a developmental plasticity gradient allowing neurodevelopmental-like cellular states stuck in plastic and stem-like or progenitor-like cellular identities, and this necessitates complex dynamics forecasting \citep{Zenil2019,Uthamacumaran2021,Uthamacumaran2022}. Predictive neural models can complement trajectory inference by learning nonlinear representations that preserve information relevant to cellular state, recurrence, or differentiation. Yet traditional feed-forward networks typically compress each cell through a fixed sequence of transformations, while Transformer models principally represent relationships through token-wise attention. These network architectures can be powerful predictors, but their hidden representations do not necessarily encode an explicit, independently unfolding internal dynamical process that can capture complex, nonlinear causal structures.

The Continuous Thought Machine, here denoted CTM or CTM$_{\mathrm{NN}}$, was introduced by Sakana AI as a neural architecture to address these conventional AI learning limits. In CTM, computation unfolds along a decoupled internal temporal dimension, or sequence of ``internal ticks,'' allowing neural activity and iterative computation to evolve independently of the temporal structure of the input data \citep{Darlow2025,SakanaCTMCode}. Its two defining learning features are neuron-level temporal processing, in which individually parameterized neuron-level models process histories of incoming pre-activations, and collective neural synchronization, computed from evolving neuron activity and used directly as a latent representation for observation and prediction. The CTM architecture therefore differs from conventional recurrent networks and Transformers not simply through recurrence, but by explicitly modeling neuron-level temporal dynamics and using synchronization between neural activities as a central representational mechanism for latent-space reconstruction. We hypothesized that these internal temporal dynamics and synchronization-based representations could be leveraged to forecast complex systems dynamics, including manifold learning of cell fate trajectories and state-transitions within single-cell ecological state-spaces.

CTM was not introduced as a systems or cell biology model, but its internal dynamical representation provides a natural computational substrate for asking whether a static transcriptomic profile can be mapped to a state or cell fate trajectory that unfolds over repeated internal iterations, akin to time-delay embedding in traditional dynamical systems theory. This internal temporal axis is relevant to cell state modelling. A transcriptomic profile is a static measurement, but it is generated by an underlying regulatory system with memory, feedback loops, and multiple timescales. CTM$_{\mathrm{NN}}$ provides a way to map each static expression profile into a synchronization-derived latent state after several internal computational iterations. The resulting vector can be interpreted as a learned representation of the regulatory configuration associated with that cell as a dynamical system. Importantly, this does not imply that CTM neuron activity is biologically identical to gene-regulatory dynamics. Rather, it provides a computational dynamical system whose internal evolution can be optimized to represent biologically defined cellular states.

We extended this architecture with reinforcement learning to construct a CTM-RL latent intervention model. A CTM autoencoder first learned to reconstruct gene expression while discriminating the cellular state of each cell. The trained CTM was then frozen, and a continuous actor-critic model learned a sequence of bounded interventions in its synchronization latent space. The reward encouraged the decoded state to approach a training-derived cellular-state prototype, preserve the correct biological class, and avoid unnecessarily large actions. Consequently, the actor learned a local policy over the CTM latent manifold rather than merely assigning a cell to a predefined category. This provides a computational representation of possible movements between state-like regions and permits a direct held-out comparison between the unmodified CTM autoencoder and the CTM followed by the learned latent policy.

A major distinction from Transformer-based analysis is that CTM synchronization represents relationships that emerge across internal thought iterations. A Transformer normally calculates token-to-token interactions through attention matrices at each layer. In contrast, CTM$_{\mathrm{NN}}$ maintains neuron-specific histories and constructs latent representations from pairwise synchronization over internal time. The current single-cell implementation does not contain a conventional stack of gene-token self-attention layers: each cell is supplied to the official CTM through a tabular adapter, and the terminal synchronization vector is used as the cell latent state. CTM-RL should therefore be presented as complementary to, rather than universally more scalable or superior to, a Transformer. Its main potential advantages are an explicit internal computational trajectory, synchronization-based representations, and the ability to inspect how predictions and latent interventions evolve across repeated internal iterations. Direct computational and predictive benchmarking against matched Transformer, recurrent, variational, optimal-transport, and RNA-velocity baselines remains future work.

Predictive importance alone is also insufficient to establish whether a gene is structurally important to the organization of a regulatory system. Gradient-based explanations can identify genes to which a trained model is sensitive, but correlated genes may receive similar importance even when they have different roles in maintaining global network organization. We therefore coupled CTM-RL representation learning with algorithmic information dynamics. Integrated gradients quantified how expression features influenced reconstruction and synchronization-latent magnitude along a path from a zero-expression baseline to the observed cell \citep{Sundararajan2017}. Decoder-output weight norms provided a complementary, model-parameter-level estimate of the genes most strongly coupled to the learned synchronization representation.

The resulting model-informed genes were used to construct correlation networks, after which their structural contribution was evaluated using the Block Decomposition Method. BDM combines local estimates derived from the Coding Theorem Method, here denoted CTM$_{\mathrm{AIT}}$, with a decomposition rule that extends algorithmic-complexity estimation to larger one- and two-dimensional objects \citep{SolerToscano2014,Zenil2018BDM}. Unlike Shannon entropy or ordinary lossless compression, BDM attempts to approximate the length of generative descriptions, including short algorithmic regularities that may not be captured by global symbol frequencies alone. For a network $G$, removal of a node or edge $e$ produces an algorithmic perturbation score
\begin{equation}
\Delta \mathrm{BDM}(e)=\mathrm{BDM}(G\setminus e)-\mathrm{BDM}(G).
\end{equation}
The magnitude $|\Delta \mathrm{BDM}(e)|$ measures how strongly the estimated description length of the network changes after that intervention. Algorithmic perturbation analysis has been proposed as a model-based framework for ranking elements by their contributions to a system's inferred generative organization and for identifying interventions that may move complex systems through information-theoretic state space \citep{Zenil2019}. In this manuscript, however, BDM is treated as an in silico hypothesis-generating score for identifying candidate network biomarkers and transition genes that may help steer or reprogram cell fate decisions.

\subsection{Rationale and scope of computational steerability}

The resulting framework joins three levels of analysis. CTM-RL learns a nonlinear synchronization manifold and a policy over that manifold. Integrated gradients and decoder weights identify the expression features used to construct it, and BDM perturbations estimate which of those features are structurally consequential within model-informed gene-association networks. BDM perturbation analysis, as a part of the algorithmic information dynamics toolkit, provides a measure of steerability or reprogrammability of cancer networks and their attractor dynamics (i.e., phenotypic transitions)\citep{Zenil2019}. This combination is intended to distinguish state markers, which are associated with a phenotype, from candidate transition regulators, which are simultaneously predictive of latent-state organization and capable of producing a large algorithmic change when perturbed. We applied the framework to H3K27M glioma cell-state and longitudinal initial-to-recurrence analyses derived from the Jessa and Sussman atlases \citep{Jessa2022,Sussman2026}. We asked four questions: (i) whether CTM-RL improved held-out prediction or reconstruction relative to CTM-AE; (ii) whether out-of-fold synchronization states formed reproducible empirical potential basins, (iii) which genes were consistently important across decoder and integrated-gradient analyses, and (iv) which of these model-informed genes and edges generated the largest BDM perturbation signatures across network thresholds.

The rationale is to connect the genes used by an interpretable latent representation to their estimated structural contributions in a model-informed network. Decoder weights and integrated gradients describe model dependence, while BDM perturbation signatures quantify changes in estimated network description length. Collectively, they nominate candidates for subsequent testing of cell-state steerability (reprogrammability) in silico\citep{Sundararajan2017,Zenil2019,Uthamacumaran2021,Uthamacumaran2022}. Here, steerability is a research objective: whether targeted interventions in silico (i.e., network perturbation analysis using BDM signatures and IG control points) could redirect a cell or cell fate trajectory towards a specified state. The present within-class RL reward does not demonstrate escape from a malignant attractor or reprogramming into a differentiated state. These analyses support computational discovery only. Neither feature attribution, density-derived basins nor node or edge removal establishes an experimentally verified causal effect on cell fate dynamics, rather they are proposed as in silico candidates for reprogramming strategies and targeted therapies.


\begin{figure}[!htbp]
\centering
\resizebox{0.98\textwidth}{!}{%
\begin{tikzpicture}[
  font=\small,
  node distance=8mm and 9mm,
  box/.style={
    draw=black,
    rounded corners=2pt,
    align=center,
    minimum height=11mm,
    text width=4.0cm,
    inner sep=5pt,
    fill=black!3
  },
  smallbox/.style={
    draw=black,
    rounded corners=2pt,
    align=center,
    minimum height=11mm,
    text width=3.35cm,
    inner sep=5pt,
    fill=black!2
  },
  arrow/.style={-{Latex[length=2.2mm]}, thick},
  group/.style={
    draw=black!55,
    rounded corners=3pt,
    dashed,
    inner sep=6pt
  }
]

\node[box] (input)
{Single-cell expression\\
Malignant-cell labels\\
Cell-stratified outer folds};

\node[box, right=of input] (ctm)
{CTM autoencoder\\
32 thought steps\\
256-dimensional synchronization latent};

\node[box, right=of ctm] (rl)
{Frozen CTM + actor--critic\\
Four bounded latent actions\\
NMF-prototype reward};

\draw[arrow] (input.east) -- (ctm.west);
\draw[arrow] (ctm.east) -- (rl.west);

\node[smallbox, below=18mm of input] (perf)
{Held-out evaluation\\
Accuracy, macro-F1, AUROC\\
MSE, MAE, $R^2$};

\node[smallbox, right=of perf] (land)
{OOF latent geometry\\
PCA, UMAP, PHATE\\
$U=-\log p$ and $-\nabla U$};

\node[smallbox, right=of land] (xai)
{Explainable AI\\
Decoder $L_2$\\
Reconstruction and latent IG};

\node[smallbox, right=of xai] (bdm)
{Model-informed networks\\
Centralities + BDM\\
Node/edge perturbations};

\draw[arrow] (ctm.south) -- (land.north);

\draw[arrow] (perf.east) -- (land.west);
\draw[arrow] (land.east) -- (xai.west);
\draw[arrow] (xai.east) -- (bdm.west);

\node[
  group,
  fit=(perf)(land)(xai)(bdm)
] (analysisgroup) {};

\node[
  font=\small\bfseries,
  below=2mm of analysisgroup.south
]
{Cross-fold consensus and cross-dataset integration};

\end{tikzpicture}%
}

\caption{Overview of scCTM-RL-AID. A CTM autoencoder learns a synchronization-derived representation from single-cell expression. A continuous actor--critic policy applies bounded actions in the frozen latent space. Held-out performance, density-derived latent landscapes, decoder and integrated-gradient XAI, and post hoc network BDM perturbations are then integrated. BDM is not included in the neural training objective, and the empirical flow $-\nabla U$ is not RNA velocity. This algorithmic pipeline serves as a complex dynamics forecasting framework for identifying in silico candidate markers steering cell fate trajectories (decision-making).}
\label{fig:workflow}
\end{figure}
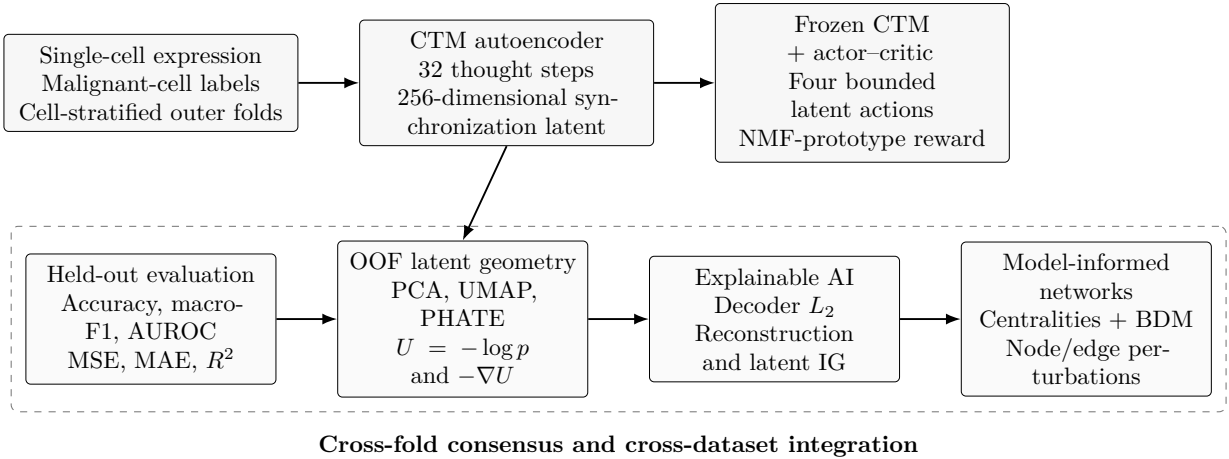

\section{Methods}

\subsection{Study design and analytical workflow}

The scCTM-RL-AID workflow was organized into two phases (Figure~\ref{fig:workflow}). Phase 1 trained and evaluated CTM autoencoder and CTM-RL models under five-fold cross-validation and generated out-of-fold latent states and feature-attribution rankings. Phase 2 pooled the completed outer folds, reconstructed Waddington-like latent landscapes, and performed gene-network and algorithmic perturbation analyses without retraining the neural models. The model-training repository is identified under Data and code availability.

The analysis included two glioma single-cell comparisons (Table~\ref{tab:datasets}): (i) GSE210567 stem-like versus differentiated malignant-cell states from the H3K27M atlas of Jessa and colleagues \citep{Jessa2022}, and (ii) H3K27M initial versus recurrent disease from the longitudinal pHGG atlas of Sussman and colleagues \citep{Sussman2026}. Where suitable metadata were available, cells were restricted to malignant populations. State labels for GSE210567 were generated from dataset annotations using terms representing neural progenitor, oligodendrocyte-progenitor, and stem-like states versus neuronal, astrocytic, mesenchymal, oligodendroglial, or otherwise differentiated states. 

\begin{table}[!htbp]
\centering
\scriptsize
\caption{Datasets, outcomes and completed cross-validation folds.}
\label{tab:datasets}
\begin{adjustbox}{max width=\textwidth}
\begin{tabular}{p{2.5cm}p{3.0cm}p{3.1cm}ccccp{3.0cm}}
\toprule
Analysis & Source & Outcome & Cells & Folds & NMF K & Interpretation \\
\midrule
GSE210567 & Jessa et al. H3K27M single-cell atlas & Stem-like (0) vs differentiated (1) & 18,490 & 5/5 & 7 & Malignant-cell state modelling \\
H3K27M initial vs recurrence & Sussman et al. longitudinal pHGG atlas & Initial (0) vs recurrence (1) & 20,000 & 5/5 & 7 & Disease-progression modelling \\
\bottomrule
\end{tabular}
\end{adjustbox}
\end{table}

\subsection{Expression preprocessing and cell sampling}

Count matrices were aligned to cell metadata using cell barcodes. Each cell was library-size normalized to $10^4$ total counts and transformed as
\begin{equation}
x_{ig}^{\prime}=\log\left(1+10^4\frac{x_{ig}}{\sum_g x_{ig}}\right),
\end{equation}
where $x_{ig}$ is the expression of gene $g$ in cell $i$. Each dataset was capped at 20,000 cells. Sampling was stratified, where possible, by patient, sample, time point, malignant status, and cellular annotation to preserve minority biological strata. A fixed random seed of 1337 was used throughout model fitting and sampling.

\subsection{NMF-derived state scaffold}

Non-negative matrix factorization (NMF) \citep{LeeSeung1999} was used to construct the training-derived state scaffold employed by the reinforcement-learning reward. Candidate ranks $K=1,\ldots,10$ were screened once for each dataset. The expression matrix was divided into a rank-screening training and validation subset, and the validation reconstruction mean-squared error was calculated for each feasible rank. The smallest rank within one standard error of the minimum validation error was selected:
\begin{equation}
K^{*}=\min\left\{K:\operatorname{MSE}_{K}\leq\operatorname{MSE}_{\min}+\operatorname{SE}_{\min}\right\}.
\end{equation}
The selected $K^{*}=7$ was fixed across all five outer folds in each dataset. To avoid fitting factor loadings on held-out cells, an independent NMF model was nevertheless fitted using only the training cells within each fold. Agglomerative Ward clustering was applied to the training-cell NMF representation, and an expression-space centroid was calculated for each NMF cluster. These centroids acted as state prototypes for CTM-RL training.

\subsection{Cross-validation design}

A stratified five-fold outer cross-validation design was used. Within each outer training partition, a validation subset was held out for neural-model selection. Model parameters, NMF factors, cluster assignments, state centroids, and RL model selection were derived without using outer-test cells. The code performs cell-level stratified cross-validation. It records patient overlap in an audit file but does not enforce patient-grouped separation. The focus is on collective intelligence and emergent behaviors at the cellular scale of transition dynamics. Further, the limited patient-level data  in the cohorts prevents such personalised risk stratification or early recurrence warning claims. The present evaluation is therefore described as five-fold cell-stratified cross-validation, not patient-held-out validation for modelling cell fate decisions in glioma ecologies as complex dynamical systems. Both retained analyses completed all five outer folds.

\subsection{Continuous Thought Machine autoencoder}

Each normalized expression vector $x_i\in\mathbb{R}^{G}$ was supplied to the official Sakana AI CTM$_{\mathrm{NN}}$ implementation through a tabular adapter. The vector was reshaped to one CTM input object containing $G$ gene features. The model used 32 internal CTM iterations, synchronization latent dimension $D=256$, CTM input dimension 128, eight synchronization heads, synapse depth two, neuron-history length eight, deep neuron-level models, no positional embedding, and no external feature-extraction backbone \citep{Darlow2025,SakanaCTMCode}. The terminal output-synchronization vector was taken as the cell representation,
\begin{equation}
z_i=\operatorname{Sync}_{\theta}(x_i)\in\mathbb{R}^{256}.
\end{equation}
A decoder mapped the synchronization state back into expression space:
\begin{equation}
\hat{x}_i=W_2\,\sigma(W_1z_i+b_1)+b_2,
\end{equation}
where $\sigma$ was a rectified linear unit. The native CTM output projector generated cellular-state logits. The CTM autoencoder was optimized using an unweighted sum of expression reconstruction and state-classification losses,
\begin{equation}
\mathcal{L}_{\mathrm{AE}}=\frac{1}{G}\lVert x_i-\hat{x}_i\rVert_2^2+\operatorname{CE}(y_i,\hat{y}_i).
\end{equation}
Training used Adam with a learning rate of $10^{-3}$, a batch size of 128, and ten epochs. The parameters producing the lowest internal-validation loss were retained.

\subsection{Reinforcement learning over CTM synchronization states}

Following autoencoder training, all CTM and decoder parameters were frozen. A Gaussian actor and scalar critic were then trained over the 256-dimensional synchronization space using an actor-critic objective \citep{Sutton1999}. The actor contained two 256-unit hidden layers and produced a mean vector $\mu(z)$ and clipped log-standard-deviation vector $\log\sigma(z)$. At each RL step,
\begin{equation}
a_t\sim\mathcal{N}\left(\mu(z_t),\operatorname{diag}[\sigma(z_t)^2]\right),
\end{equation}
and the latent state was updated according to
\begin{equation}
z_{t+1}=z_t+\eta a_t,\qquad \eta=0.1.
\end{equation}
Four latent-action steps were applied to each cell. The terminal latent state was decoded into an expression profile and passed through the frozen CTM classification head.

For a cell assigned to NMF cluster $c_i$, with expression centroid $\bar{x}_{c_i}$, the reward was
\begin{equation}
r_t=-\lambda_{\mathrm{rec}}\lVert\hat{x}_{t+1}-\bar{x}_{c_i}\rVert_2^2+\lambda_{\mathrm{cls}}\log p_{\theta}(y_i\mid z_{t+1})-\lambda_{\mathrm{act}}\lVert a_t\rVert_2^2,
\end{equation}
with $\lambda_{\mathrm{rec}}=1$, $\lambda_{\mathrm{cls}}=1$, and $\lambda_{\mathrm{act}}=10^{-3}$. Thus, the policy was rewarded for moving the decoded cell toward its training-derived state prototype, retaining the correct biological label, and minimizing the size of the latent intervention. Discounted returns were calculated with $\gamma=0.99$. The combined actor-critic objective was
\begin{equation}
\mathcal{L}_{\mathrm{RL}}=\mathcal{L}_{\mathrm{policy}}+0.5\mathcal{L}_{\mathrm{value}}-10^{-3}\mathcal{H}(\pi),
\end{equation}
where $\mathcal{H}(\pi)$ is the Gaussian policy entropy. The actor and critic were trained for eight epochs using Adam with a learning rate of $3\times10^{-4}$. The model with the highest internal-validation reward was retained. Outer-test inference used the deterministic actor mean rather than sampled actions.

This construction is described as a latent intervention policy or computational cell-state transition policy. Because no lineage-paired cell transitions were supplied during training, it is not an experimentally supervised model of physical transition velocity. Our computational and systems medicine approaches are dynamical systems alternatives to traditional statistically bound trajectory inference algorithms. The RL rollout also does not create an independent biological time variable. It produces a scaled, action-penalized computational displacement in the learned synchronization space.

\subsection{Held-out model evaluation}

Two models were evaluated on every completed outer test fold: CTM-AE and CTM-AE+RL. The second consisted of the same frozen CTM autoencoder followed by the deterministic RL actor rollout. The code reports accuracy, macro-F1, AUROC, expression-space MSE, MAE, and $R^2$. Values were summarized as mean, standard deviation, and Student-$t$ 95\% confidence interval across completed folds. Paired fold-level RL-minus-AE differences were calculated descriptively. 

\subsection{Explainable AI and gene-feature attribution}

Three complementary feature-importance measures were retained. First, for each gene $g$, decoder importance was defined as the $L_2$ norm of the weights mapping the terminal decoder hidden state to that gene,
\begin{equation}
I^{\mathrm{decoder}}_g=\lVert W^{\mathrm{out}}_{g,:}\rVert_2.
\end{equation}
This measure identifies genes whose reconstruction depends strongly on the learned synchronization representation.

Second, integrated gradients were calculated from an all-zero expression baseline $x'$ to each observed expression vector $x$ \citep{Sundararajan2017}:
\begin{equation}
\operatorname{IG}_g(x)=(x_g-x'_g)\int_0^1\frac{\partial F(x'+\alpha(x-x'))}{\partial x_g}\,d\alpha.
\end{equation}
The reconstruction target was the negative reconstruction error,
\begin{equation}
F_{\mathrm{rec}}(x)=-\frac{1}{G}\lVert\hat{x}-x\rVert_2^2,
\end{equation}
so high absolute attribution indicated genes influencing faithful reconstruction. Third, latent attribution used the mean squared synchronization magnitude,
\begin{equation}
F_{\mathrm{latent}}(x)=\frac{1}{D}\sum_{d=1}^D z_d(x)^2.
\end{equation}
This quantified how individual genes contributed to formation of the synchronization latent state. Integrated gradients were calculated separately for CTM-AE and CTM-AE+RL using a trapezoidal approximation with 24 integration steps and up to 200 held-out cells per fold. Absolute attributions were averaged across cells. Claims are limited to decoder, reconstruction, and latent-state importance.

\subsection{Cross-fold feature consensus and Explainable AI (XAI)}

The 100 highest-ranked genes from each completed fold were pooled separately for decoder $L_2$ importance, reconstruction integrated gradients, and latent integrated gradients. Scores were converted to within-file normalized values and aggregated according to fold recurrence, model-source recurrence, mean normalized importance, and mean rank. A union set was then constructed by prioritizing genes found across multiple feature-importance classes. This generated four model-informed gene sets per dataset: decoder, reconstruction IG, latent IG, and their union. 
\subsection{Out-of-fold latent Waddington-like landscapes}

Only latent vectors generated for outer-test cells by CTM-AE+RL were used to construct the landscapes. Completed outer folds were combined so that each cell contributed at most one out-of-fold latent representation. Latent dimensions with negligible variance were removed, remaining dimensions were standardized, and representations wider than 64 dimensions were reduced to 64 principal components. Up to 5,000 cells were sampled while preserving class proportions. Three separate three-dimensional embeddings were generated using PCA, UMAP \citep{McInnes2018}, and PHATE \citep{Moon2019}.

For each embedding, a Gaussian kernel-density estimate $p(s)$ was calculated from the first two embedding coordinates. The empirical potential was defined as
\begin{equation}
U(s)=-\log[p(s)+\epsilon],
\end{equation}
where $\epsilon$ prevented numerical divergence in regions of negligible estimated density. A gradient field was calculated as
\begin{equation}
F_{\mathrm{pot}}(s)=-\nabla U(s).
\end{equation}
Low-potential, high-density regions were interpreted as candidate state basins, whereas sparsely populated high-potential regions represented barriers or transition-like regions. For quantitative description, we calculated class silhouette coefficient, normalized centroid separation, and the AUROC with which $-U$ distinguished class 1 from class 0. These are descriptive embedding-level summaries, not proof of a biological potential function.

The code also generated a high-$U$-to-low-$U$ rank coordinate. This is called a potential-ranked landscape coordinate rather than diffusion pseudotime because it is derived from density rank and does not use a lineage root, transition matrix, measured temporal direction, or RNA-velocity model.

\subsection{Model-informed gene-network reconstruction}

For each pooled feature set, expression was loaded for the corresponding genes using a common cell matrix. Pairwise Spearman correlations were calculated, and $P$ values were corrected using the Benjamini-Hochberg procedure \citep{Benjamini1995}. Undirected networks were retained at $|\rho|\geq0.30$, $0.50$, or $0.70$ with $q\leq0.05$. Each thresholded network was represented as a binary symmetric adjacency matrix. Weighted degree, strength, betweenness, closeness, eigenvector centrality, PageRank, clustering coefficient, and core number were calculated. Leiden community detection \citep{Traag2019} was applied to identify network modules.

\subsection{BDM and algorithmic perturbation signatures}

Two-dimensional BDM was calculated using \texttt{pyBDM} with a binary alphabet. The package decomposes larger one- and two-dimensional arrays into blocks whose local complexities are estimated from precomputed CTM$_{\mathrm{AIT}}$ values \citep{SolerToscano2014,Zenil2018BDM}. For each thresholded adjacency matrix $A$, baseline complexity was
\begin{equation}
K_{\mathrm{BDM}}(A)=\mathrm{BDM}_{2D}(A).
\end{equation}
For node $g$, the corresponding row and column were set to zero while preserving the matrix dimensions:
\begin{equation}
\Delta_g=\mathrm{BDM}_{2D}(A^{(-g)})-\mathrm{BDM}_{2D}(A).
\end{equation}
For an edge $e=(g_i,g_j)$,
\begin{equation}
\Delta_e=\mathrm{BDM}_{2D}(A^{(-e)})-\mathrm{BDM}_{2D}(A).
\end{equation}
Genes and edges were ranked primarily by the mean absolute perturbation across three correlation thresholds, $\mathcal{T}=\{0.30,0.50,0.70\}$,
\begin{equation}
S_g^{\mathrm{BDM}}=\frac{1}{|\mathcal{T}|}\sum_{\tau\in\mathcal{T}}|\Delta_g^{(\tau)}|.
\end{equation}
Signed values were retained to indicate whether a perturbation increased or decreased estimated algorithmic description length, while absolute values quantified intervention magnitude. Edge perturbation analysis was capped at the 1,000 strongest retained edges when a network exceeded that number.

To assess whether BDM captured information distinct from statistical disorder or ordinary compressibility, equivalent perturbation shifts were calculated for binary Shannon entropy, GZIP size, and actual LZW emitted-code count. Spearman correlations between their absolute shifts and BDM shifts were calculated at each threshold. Strong concordance was treated as evidence that network sparsity or simple compressibility explained part of the ranking, rather than as independent confirmation.

\subsection{Integration of latent and algorithmic signatures}

The integration was performed at the interpretation and ranking levels:
\begin{equation}
\text{expression}\rightarrow\text{CTM synchronization latent}\rightarrow\text{XAI gene set}\rightarrow\text{gene network}\rightarrow\Delta\mathrm{BDM}.
\end{equation}
Genes were considered strongest candidate transition regulators when they showed recurrent selection across completed outer folds, high reconstruction or latent integrated-gradient importance, strong decoder contribution, central placement within a model-informed gene network, a large and threshold-stable BDM node-removal shift, and reproducibility across datasets or related comparisons. A transparent cross-dataset reciprocal-rank summary was calculated as $1/(r_{\mathrm{XAI}}+1)+1/(r_{\mathrm{BDM}}+1)$ within each dataset, assigning rank 101 to genes absent from a top-100 list. This score is used for prioritization only and was not optimized during model training.

BDM scores were not included in the CTM-RL reward and did not alter neural-network training. The analysis is therefore a post hoc coupling of representation learning and algorithmic perturbation analysis as network intervention ranking, not an end-to-end BDM-regularized neural model. Likewise, ``causal'' refers to model-based in silico perturbation ranking in the algorithmic-information sense \citep{Zenil2019}. Experimental perturbation, longitudinal lineage tracing, and intervention-response validation are required before biological causal claims can be made.
\subsection{Exploratory functional enrichment}
\label{sec:enrichment_methods}

Functional overrepresentation was tested separately for the top-30 cross-fold consensus decoder, reconstruction-IG, latent-IG, XAI-union and BDM-node signatures in each dataset. HGNC-approved symbols and unambiguous aliases were used \citep{Seal2023}. GO Biological Process 2025 and Reactome Pathways 2024 annotations were downloaded from Enrichr \citep{Kuleshov2016}. Statistical tests were calculated locally in Jupyter notebook using Python 3.12.13, NumPy 2.4.4, pandas 2.3.3 and SciPy 1.17.1, with Matplotlib 3.10.8 for plotting. Each library's mapped annotated human genes formed the background (GO: 14,653; Reactome: 11,375). Terms with 3--500 background genes were tested using a one-sided hypergeometric probability, $P=\Pr(X\geq k)$ for $X\sim\mathrm{Hypergeom}(N,K,n)$, where $N$ is background size, $K$ term size, $n$ the annotated query size and $k$ the overlap. Benjamini--Hochberg correction covered all 73,790 tests across ten signatures and both libraries, including zero-overlap terms; significance required $q<0.05$. Plots retained significant terms with at least two overlapping genes and reduced redundant labels without changing the tests. Top BDM-edge endpoints were annotated descriptively without enrichment P values. 

\section{Results}

\subsection{Dataset completion and optimization behaviour}

Both retained analyses selected NMF rank $K=7$ by the prespecified one-standard-error rule (Supplementary Figure~\ref{fig:s1}). GSE210567 included 18,490 cells and the initial-versus-recurrence analysis used the 20,000-cell cap. Both analyses completed all five outer folds.

\begin{figure}[!htbp]
\centering
\begin{subfigure}[t]{0.485\textwidth}\caption{Classification performance}\includegraphics[width=\linewidth]{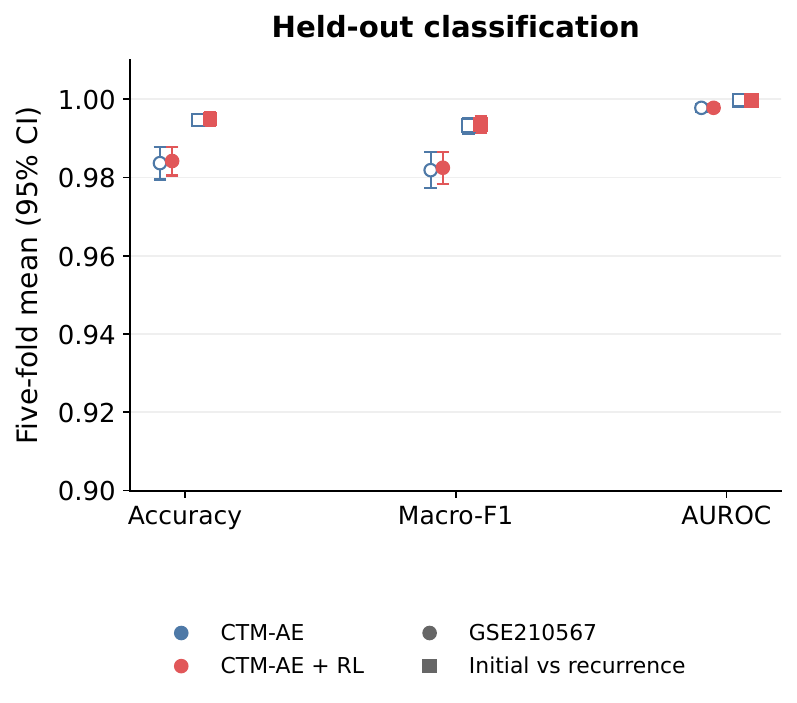}\end{subfigure}\hfill
\begin{subfigure}[t]{0.485\textwidth}\caption{Reconstruction performance}\includegraphics[width=\linewidth]{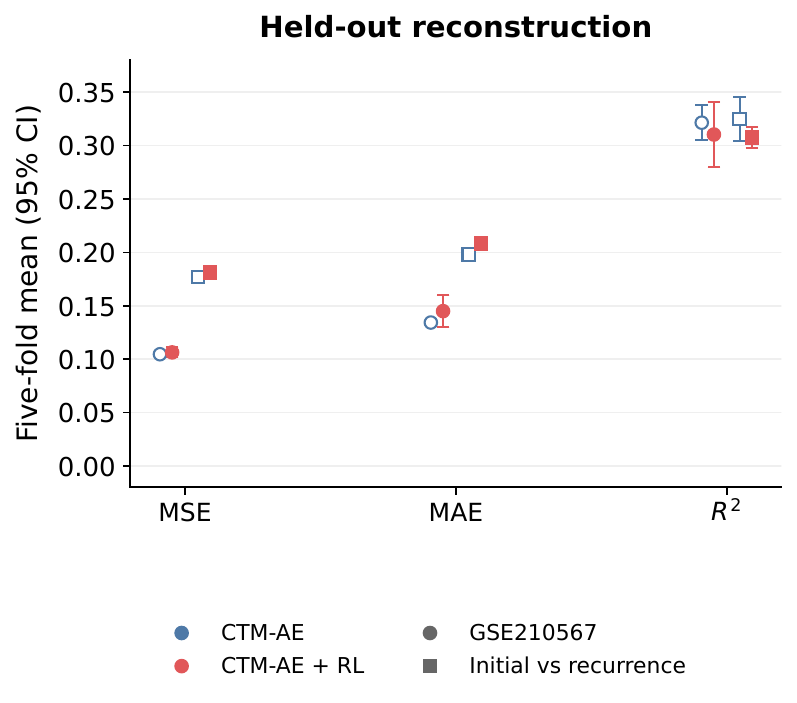}\end{subfigure}

\vspace{3mm}
\begin{subfigure}[t]{0.485\textwidth}\caption{CTM-AE optimization}\includegraphics[width=\linewidth]{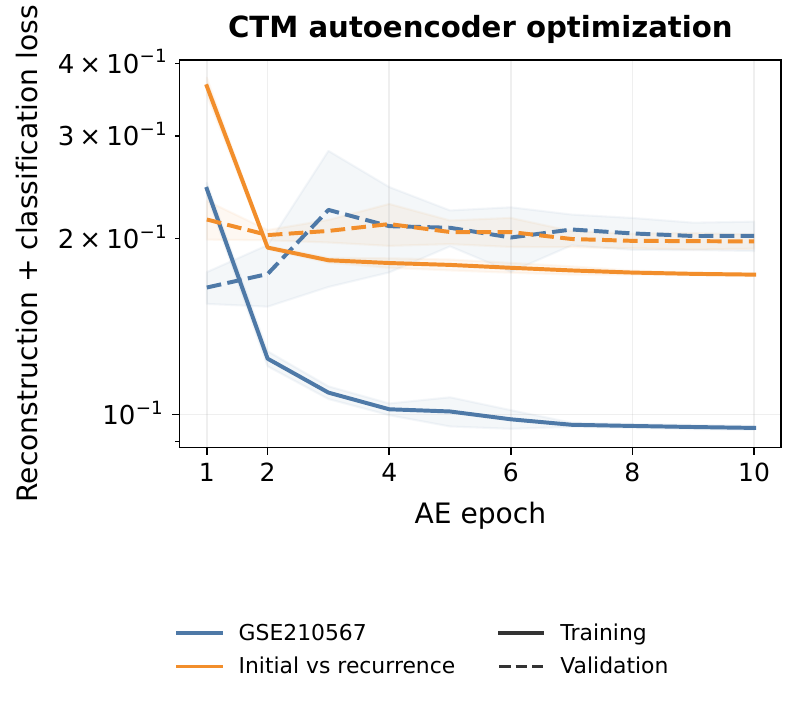}\end{subfigure}\hfill
\begin{subfigure}[t]{0.485\textwidth}\caption{Actor--critic optimization}\includegraphics[width=\linewidth]{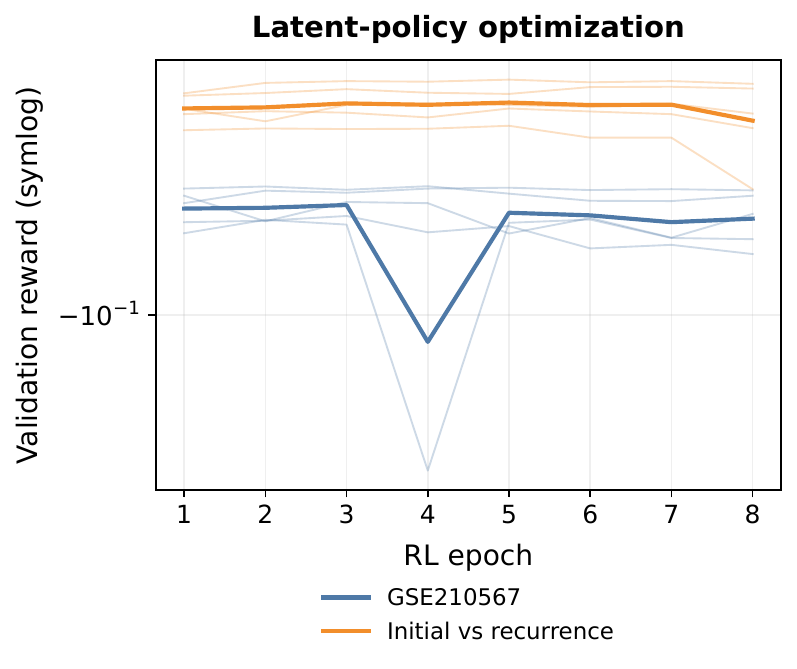}\end{subfigure}
\caption{CTM-AE versus CTM-RL benchmarking and optimization across the two retained five-fold analyses. (A) Accuracy, macro-F1, and AUROC. (B) Expression reconstruction MSE, MAE, and $R^2$. Markers represent five-fold means and error bars represent Student-$t$ 95\% confidence intervals. (C) Autoencoder training and validation losses. (D) RL validation reward trajectories. Thin curves show individual folds and thick curves show their epoch-wise means.}
\label{fig:benchmark}
\end{figure}

CTM-AE training loss decreased in both analyses, but validation behaviour differed (Figure~\ref{fig:benchmark}C). In GSE210567, the mean training loss converged near 0.095 and validation loss remained in the approximate 0.15--0.25 range. In the initial-versus-recurrence model, training loss continued to decline while validation loss often rose after early epochs, indicating overfitting despite selection of the best validation checkpoint. RL validation reward was comparatively stable in GSE210567 and variable but finite in initial versus recurrence (Figure~\ref{fig:benchmark}D).

\subsection{CTM-RL preserved classification but did not improve reconstruction from CTM-AE}

Performance across completed outer folds is shown in Table~\ref{tab:performance} and Figure~\ref{fig:benchmark}. In GSE210567, CTM-AE achieved accuracy $0.9837\pm0.0034$, macro-F1 $0.9819\pm0.0037$, and AUROC $0.9978\pm0.0008$. CTM-RL yielded accuracy $0.9842\pm0.0030$, macro-F1 $0.9825\pm0.0033$, and AUROC $0.9978\pm0.0008$. The fold-paired RL-minus-AE differences were small: $+0.00054$ for accuracy, $+0.00062$ for F1, and $+0.000005$ for AUROC, with all corresponding 95\% confidence intervals spanning zero. Given the simplicity of the binary classification problems, these metric values reflect the ease at which the models learn the discrimination of cell states rather than evidence of data leakage or overfitting (which has been carefully tested for in the five-fold experimental design with validation datasets).

The initial-versus-recurrence comparison was almost perfectly separable. CTM-AE achieved accuracy $0.9947\pm0.0013$, macro-F1 $0.9932\pm0.0016$, and AUROC $0.99974\pm0.00008$. CTM-RL achieved accuracy $0.9950\pm0.0014$, macro-F1 $0.9936\pm0.0018$, and AUROC $0.99974\pm0.00008$. The mean AUROC difference was approximately $1\times10^{-6}$. Thus, the RL rollout did not add meaningful discriminative information beyond the CTM autoencoder for this task.

Reconstruction showed a consistent disadvantage for the latent actor. In GSE210567, AE MSE was $0.1047\pm0.0020$ compared with $0.1064\pm0.0038$ after RL. MAE increased from $0.1343\pm0.0025$ to $0.1450\pm0.0121$, and $R^2$ decreased from $0.3214\pm0.0133$ to $0.3103\pm0.0248$. In initial versus recurrence, MSE increased from $0.1769\pm0.0044$ to $0.1814\pm0.0021$, MAE from $0.1982\pm0.0028$ to $0.2081\pm0.0032$, and $R^2$ decreased from $0.3249\pm0.0169$ to $0.3078\pm0.0080$. The fold-level 95\% confidence intervals for RL-minus-AE MSE, MAE, and $R^2$ excluded zero in the initial-versus-recurrence task. These findings establish that the present reward did not optimize global reconstruction fidelity and that CTM-RL should not be presented as a superior autoencoder.

\begin{table}[!htbp]
\centering
\scriptsize
\caption{Outer-fold performance. Values are mean $\pm$ SD across completed folds.}
\label{tab:performance}
\begin{adjustbox}{max width=\textwidth}
\begin{tabular}{p{2.5cm}lccccccc}
\toprule
Dataset & Model & Folds & Accuracy & Macro-F1 & AUROC & MSE & MAE & $R^2$ \\
\midrule
GSE210567 & CTM-AE & 5 & 0.9837 $\pm$ 0.0034 & 0.9819 $\pm$ 0.0037 & 0.9978 $\pm$ 0.0008 & 0.1047 $\pm$ 0.0020 & 0.1343 $\pm$ 0.0025 & 0.3214 $\pm$ 0.0133 \\
GSE210567 & CTM-RL & 5 & 0.9842 $\pm$ 0.0030 & 0.9825 $\pm$ 0.0033 & 0.9978 $\pm$ 0.0007 & 0.1064 $\pm$ 0.0038 & 0.1450 $\pm$ 0.0121 & 0.3103 $\pm$ 0.0248 \\
Initial vs recurrence & CTM-AE & 5 & 0.9947 $\pm$ 0.0013 & 0.9932 $\pm$ 0.0016 & 0.9997 $\pm$ 0.0001 & 0.1769 $\pm$ 0.0044 & 0.1982 $\pm$ 0.0028 & 0.3249 $\pm$ 0.0169 \\
Initial vs recurrence & CTM-RL & 5 & 0.9950 $\pm$ 0.0014 & 0.9936 $\pm$ 0.0018 & 0.9997 $\pm$ 0.0001 & 0.1814 $\pm$ 0.0021 & 0.2081 $\pm$ 0.0032 & 0.3078 $\pm$ 0.0080 \\
\bottomrule
\end{tabular}
\end{adjustbox}
\end{table}

\subsection{Out-of-fold synchronization states formed empirical potential attractor basins}

\begin{figure}[!htbp]
\centering
\begin{subfigure}[t]{0.485\textwidth}\caption{GSE210567 cell-state landscape}\includegraphics[width=\linewidth]{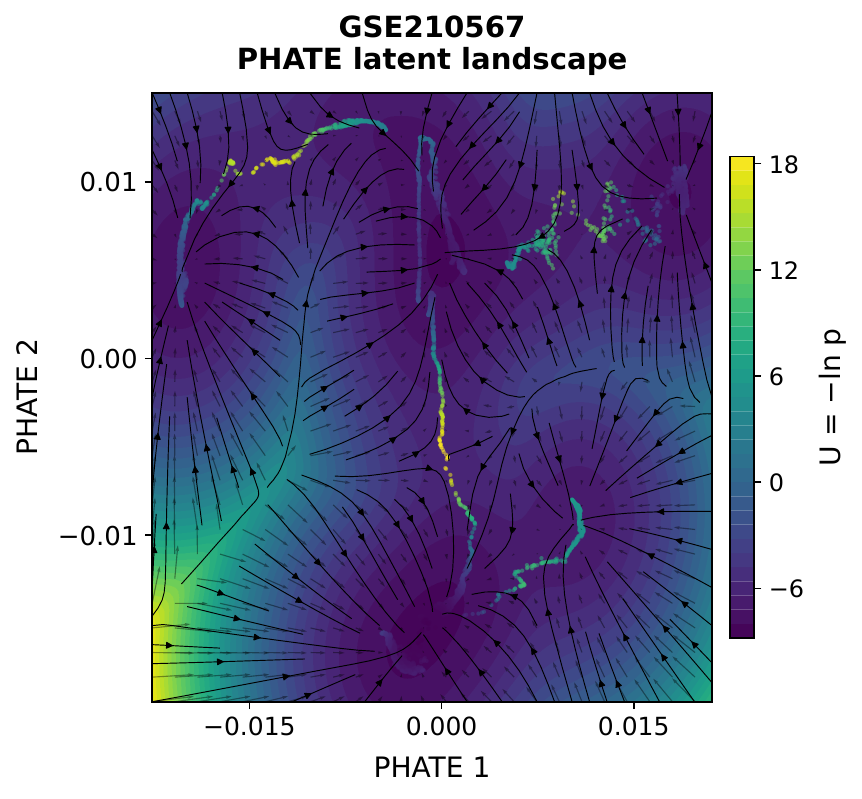}\end{subfigure}\hfill
\begin{subfigure}[t]{0.485\textwidth}\caption{Initial-versus-recurrence landscape}\includegraphics[width=\linewidth]{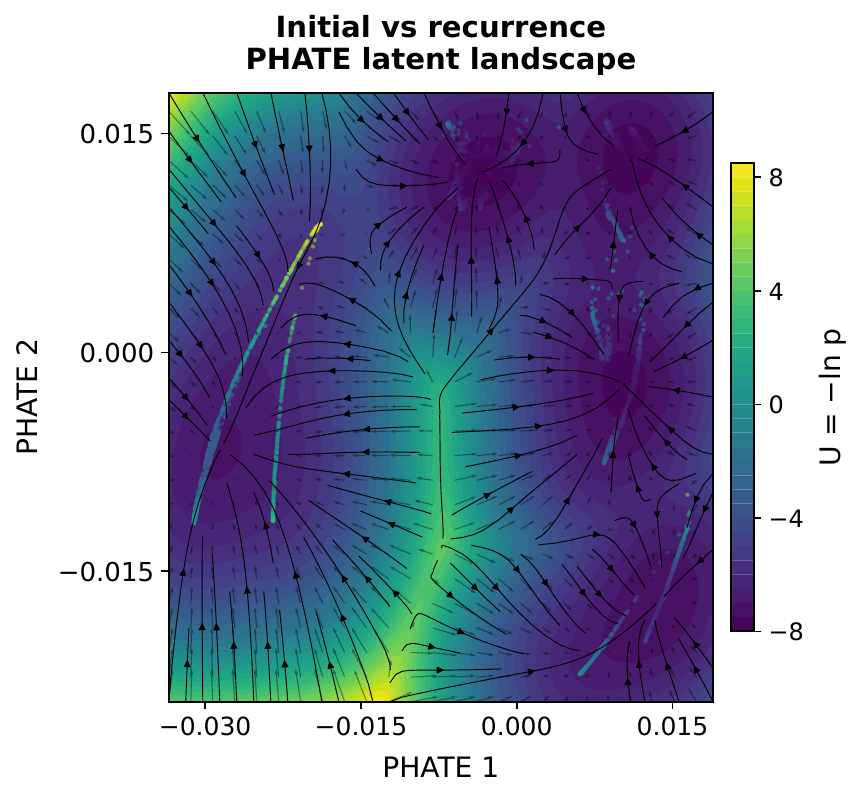}\end{subfigure}

\vspace{3mm}
\begin{subfigure}[t]{0.62\textwidth}\caption{Landscape quantification}\includegraphics[width=\linewidth]{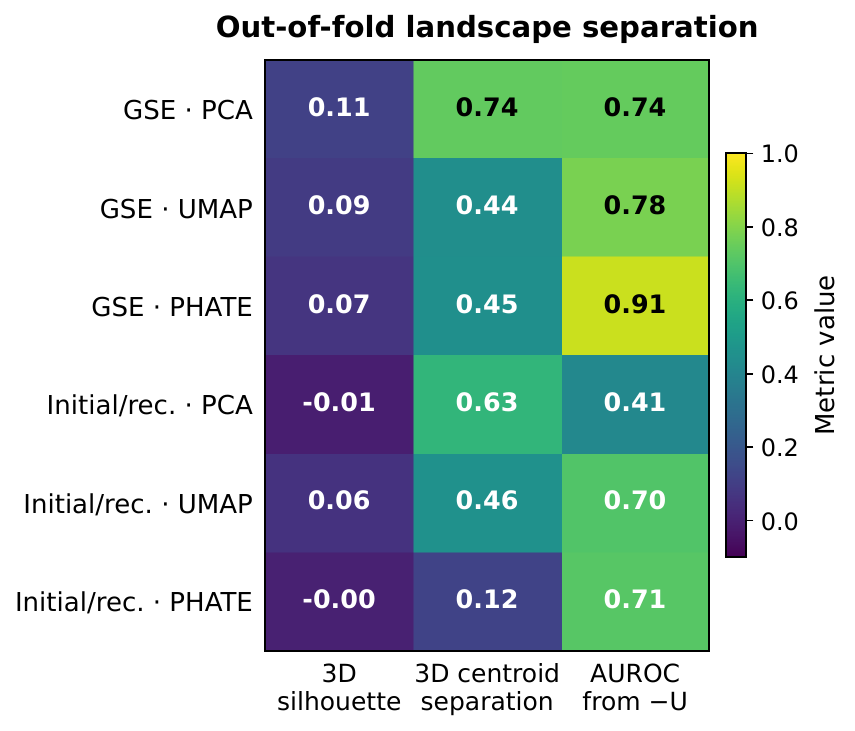}\end{subfigure}
\caption{Out-of-fold CTM-RL synchronization landscapes. (A--B) PHATE embeddings with the empirical potential $U=-\log[p(s)+\epsilon]$ and density-gradient streamlines $-\nabla U$ for the two retained analyses. These streamlines are visual summaries of the kernel-density surface, not observed trajectories or RNA velocity. (C) Embedding-level silhouette coefficients, normalized centroid separation, and class discrimination by $-U$. Full PCA, UMAP, and PHATE two- and three-dimensional outputs are retained in the Supplementary Information.}
\label{fig:landscapes}
\end{figure}

Despite minimal changes in classification, deterministic CTM-RL rollouts produced structured out-of-fold latent distributions (Figure~\ref{fig:landscapes}). In GSE210567, PCA gave the strongest geometric class separation, with a two-dimensional silhouette coefficient of 0.203 and normalized two-dimensional centroid separation of 0.888. PHATE had lower geometric separation but the largest potential-density discrimination: the mean PHATE potential was $-7.194$ for class 0 and $-7.956$ for class 1, and $-U$ distinguished the differentiated class with AUROC 0.911. The corresponding values were 0.739 for PCA and 0.781 for UMAP.

Initial versus recurrence displayed near-complete predictive separation in the CTM output head but only modest separation in unsupervised low-dimensional geometry. PCA had a negative three-dimensional silhouette coefficient ($-0.009$) despite normalized centroid separation of 0.626. UMAP produced a three-dimensional silhouette coefficient of 0.062, and PHATE remained near zero. The recurrent class nevertheless occupied lower-potential regions on average in UMAP and PHATE: PHATE mean $U$ was $-7.052$ for initial and $-7.345$ for recurrence, with $-U$ AUROC 0.712. This difference between classification and low-dimensional separation indicates that the discriminative information was distributed across the full synchronization representation, as a collective learning, and was not fully preserved in a two- or three-dimensional projection.

The streamlines in Figure~\ref{fig:landscapes} represent $-\nabla U$ on the two-dimensional kernel-density surface. They are not to be interpreted as inferred RNA velocities or observed cell trajectories.

\subsection{Decoder and integrated-gradient XAI recovered distinct but convergent cell fate signatures}

\begin{figure}[!htbp]
\centering
\begin{subfigure}[t]{0.485\textwidth}\caption{Decoder $L_2$ importance}\includegraphics[width=\linewidth]{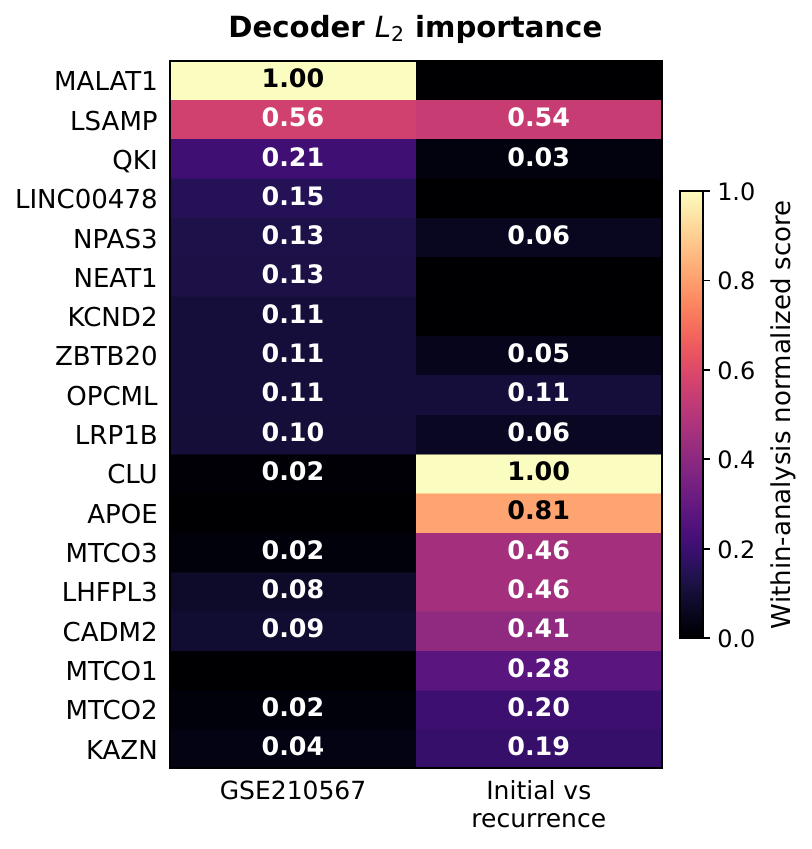}\end{subfigure}\hfill
\begin{subfigure}[t]{0.485\textwidth}\caption{Reconstruction integrated gradients}\includegraphics[width=\linewidth]{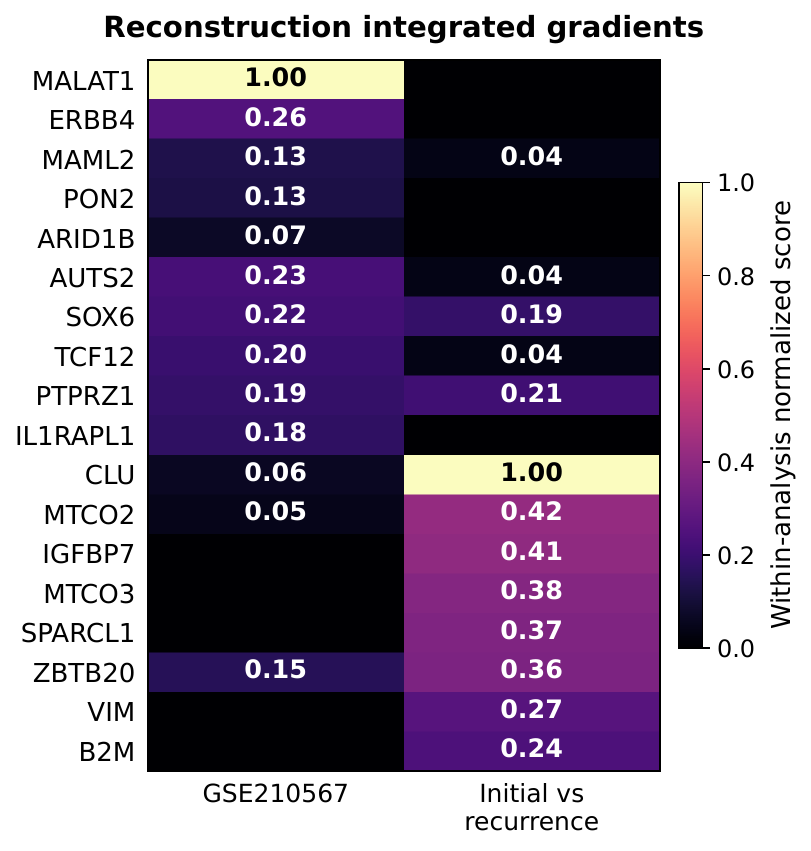}\end{subfigure}

\vspace{3mm}
\begin{subfigure}[t]{0.485\textwidth}\caption{Latent integrated gradients}\includegraphics[width=\linewidth]{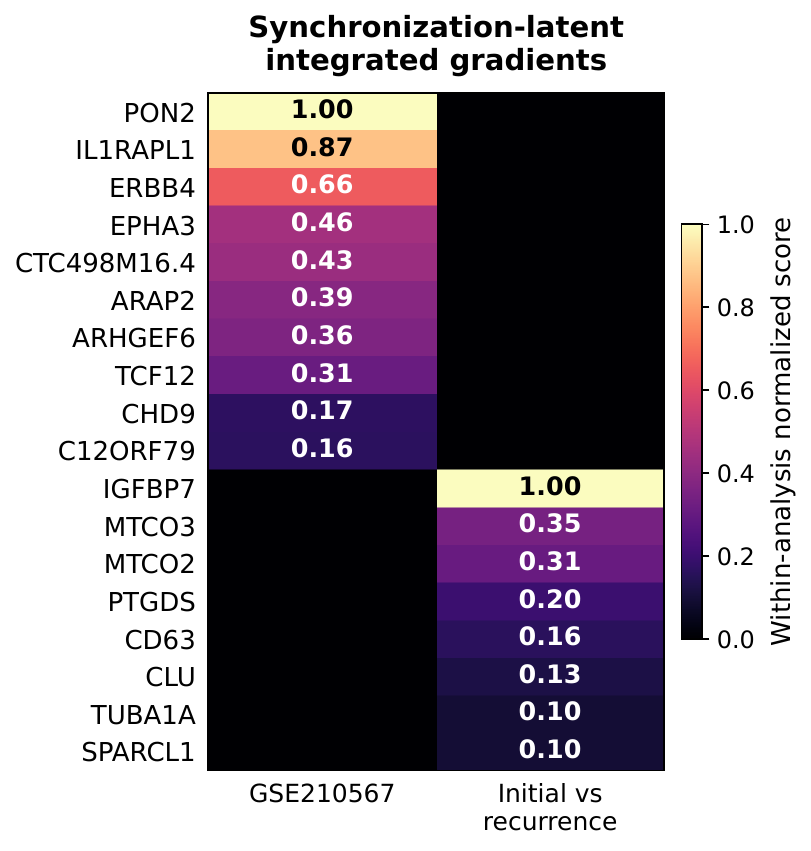}\end{subfigure}\hfill
\begin{subfigure}[t]{0.485\textwidth}\caption{Global cross-dataset consensus}\includegraphics[width=\linewidth]{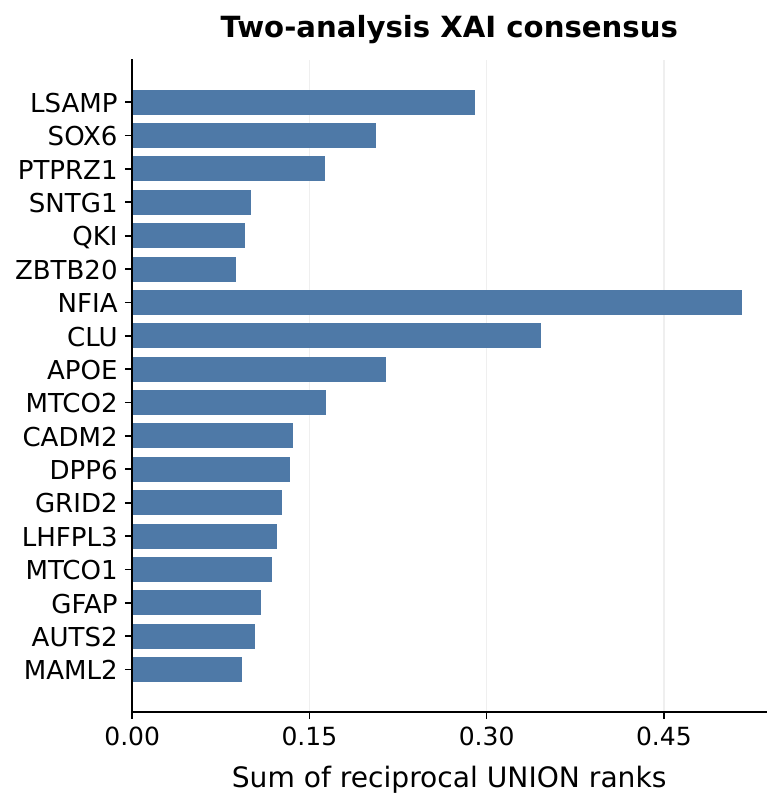}\end{subfigure}
\caption{Cross-fold explainable-AI signatures across the two retained analyses. (A) Decoder output-weight $L_2$ importance. (B) Absolute integrated gradients for the negative reconstruction-error target. (C) Absolute integrated gradients for synchronization-latent magnitude. (D) Two-analysis consensus from the unchanged per-analysis UNION ranks. The three methods probe complementary model components. Absolute attribution quantifies predictive dependence but not class-specific expression direction.}
\label{fig:xai}
\end{figure}

The three attribution families emphasized complementary aspects of the CTM representation (Figure~\ref{fig:xai}; Table~\ref{tab:xai}). Decoder $L_2$ scores highlighted genes strongly coupled to reconstruction weights, whereas reconstruction IG and latent IG measured input sensitivity of reconstruction fidelity and synchronization magnitude, respectively. Absolute IG scores do not encode whether a gene was higher in one biological class, so the reported genes are predictive dependencies rather than directional differential-expression markers. These XAI metrics provide rich insights into the neural network models' learning patterns and collective dynamics in the discrimination task.

In GSE210567, decoder importance prioritized \textit{MALAT1, LSAMP, QKI, LINC00478, NPAS3, NEAT1, KCND2, ZBTB20, OPCML}, and \textit{LRP1B}. Reconstruction IG prioritized \textit{MALAT1, ERBB4, MAML2, PON2, ARID1B, AUTS2, SOX6, TCF12, PTPRZ1}, and \textit{IL1RAPL1}. Latent IG prioritized \textit{PON2, IL1RAPL1, ERBB4, EPHA3, CTC498M16.4, ARAP2, ARHGEF6, TCF12, CHD9}, and \textit{C12ORF79}. The union consensus integrated these signals and ranked \textit{NFIA, IL1RAPL1, ERBB4, BAI3/ADGRB3, SOX6, LPHN3/ADGRL3, PTPRZ1, DPP6, GRID2}, and \textit{AUTS2} as its top ten.

The initial-versus-recurrence task produced a progression-associated XAI signature spanning extracellular, stress, mitochondrial, and neural-identity components. Decoder $L_2$ prioritized \textit{CLU, APOE, LSAMP, MT-CO3, LHFPL3, CADM2, MT-CO1, MT-CO2, KAZN}, and \textit{SNTG1}. Reconstruction IG prioritized \textit{CLU, MT-CO2, IGFBP7, MT-CO3, SPARCL1, ZBTB20, VIM, B2M, SOX6}, and \textit{CD63}. Latent IG prioritized \textit{IGFBP7, MT-CO3, MT-CO2, PTGDS, CD63, CLU, TUBA1A, SPARCL1, MT-ND3}, and \textit{MT-ND4}. The integrated union ranked \textit{IGFBP7, CLU, LSAMP, APOE, MT-CO3, MT-CO2, CADM2, LHFPL3, MT-CO1}, and \textit{SPARCL1}.

Across the two retained analyses, top-30 XAI recurrence was shared by \textit{LSAMP, SOX6, PTPRZ1, SNTG1, QKI}, and \textit{ZBTB20}. The two-analysis consensus summarizes recurrence of the unchanged per-analysis UNION rankings and is not a pooled expression effect size.

\begin{table}[!htbp]
\centering
\scriptsize
\caption{Top cross-fold predictive signatures. IG denotes integrated gradients.}
\label{tab:xai}
\begin{adjustbox}{max width=\textwidth}
\begin{tabular}{p{2.7cm}p{2.7cm}p{10.0cm}}
\toprule
Dataset & Analysis & Top ten consensus genes \\
\midrule
GSE210567 & Decoder $L_2$ & MALAT1, LSAMP, QKI, LINC00478, NPAS3, NEAT1, KCND2, ZBTB20, OPCML, LRP1B \\
GSE210567 & Reconstruction IG & MALAT1, ERBB4, MAML2, PON2, ARID1B, AUTS2, SOX6, TCF12, PTPRZ1, IL1RAPL1 \\
GSE210567 & Latent IG & PON2, IL1RAPL1, ERBB4, EPHA3, CTC498M16.4, ARAP2, ARHGEF6, TCF12, CHD9, C12ORF79 \\
GSE210567 & Integrated union & NFIA, IL1RAPL1, ERBB4, BAI3, SOX6, LPHN3, PTPRZ1, DPP6, GRID2, AUTS2 \\
Initial vs recurrence & Decoder $L_2$ & CLU, APOE, LSAMP, MTCO3, LHFPL3, CADM2, MTCO1, MTCO2, KAZN, SNTG1 \\
Initial vs recurrence & Reconstruction IG & CLU, MTCO2, IGFBP7, MTCO3, SPARCL1, ZBTB20, VIM, B2M, SOX6, CD63 \\
Initial vs recurrence & Latent IG & IGFBP7, MTCO3, MTCO2, PTGDS, CD63, CLU, TUBA1A, SPARCL1, MTND3, MTND4 \\
Initial vs recurrence & Integrated union & IGFBP7, CLU, LSAMP, APOE, MTCO3, MTCO2, CADM2, LHFPL3, MTCO1, SPARCL1 \\
\bottomrule
\end{tabular}
\end{adjustbox}
\end{table}

\subsection{BDM perturbations separated structural importance from conventional centrality}

\begin{figure}[!htbp]
\centering
\begin{subfigure}[t]{0.485\textwidth}\caption{GSE210567 BDM perturbations}\includegraphics[width=\linewidth]{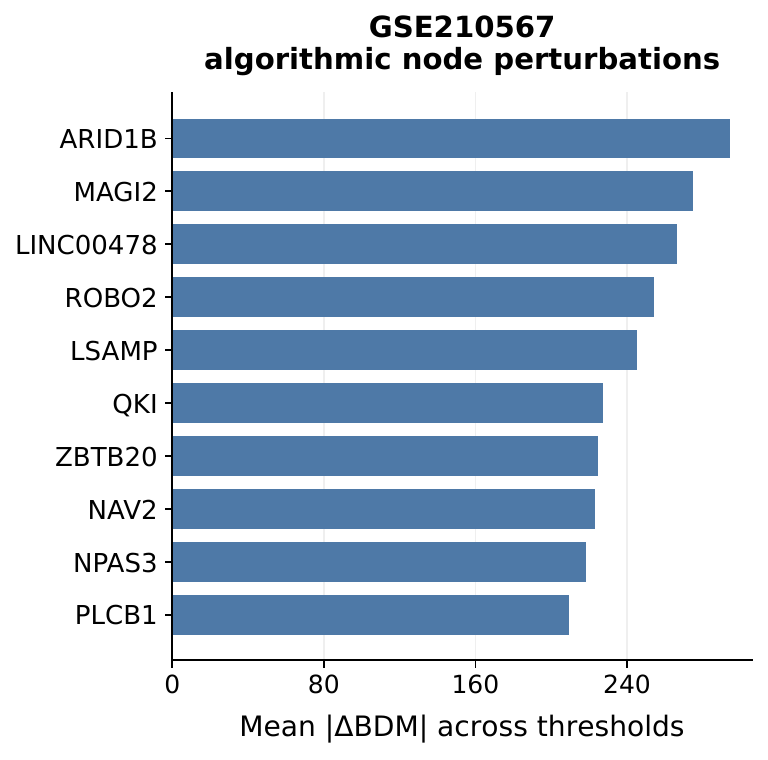}\end{subfigure}\hfill
\begin{subfigure}[t]{0.485\textwidth}\caption{Initial-versus-recurrence perturbations}\includegraphics[width=\linewidth]{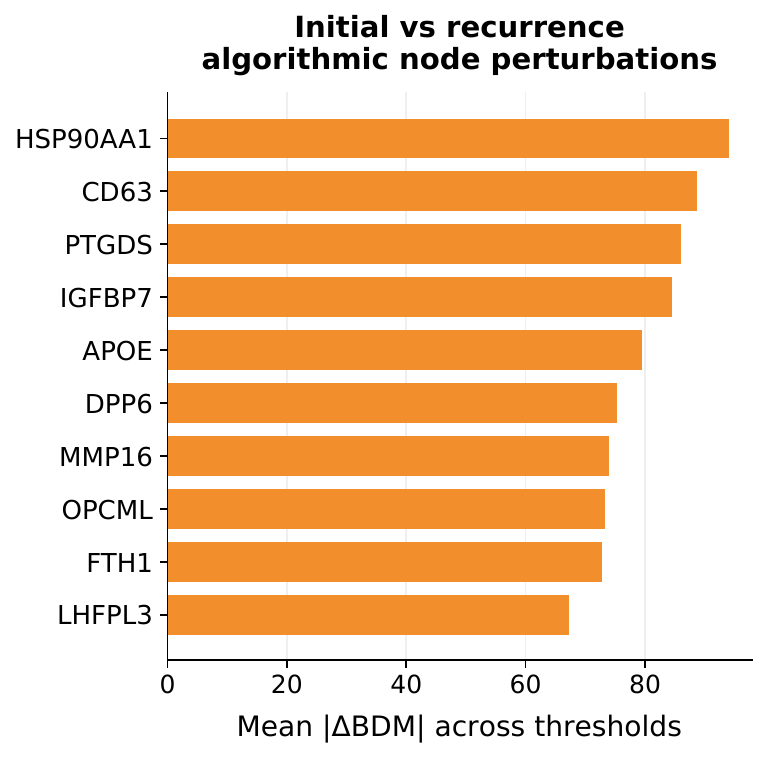}\end{subfigure}

\vspace{3mm}

\begin{subfigure}[t]{0.62\textwidth}\caption{BDM versus degree}\includegraphics[width=\linewidth]{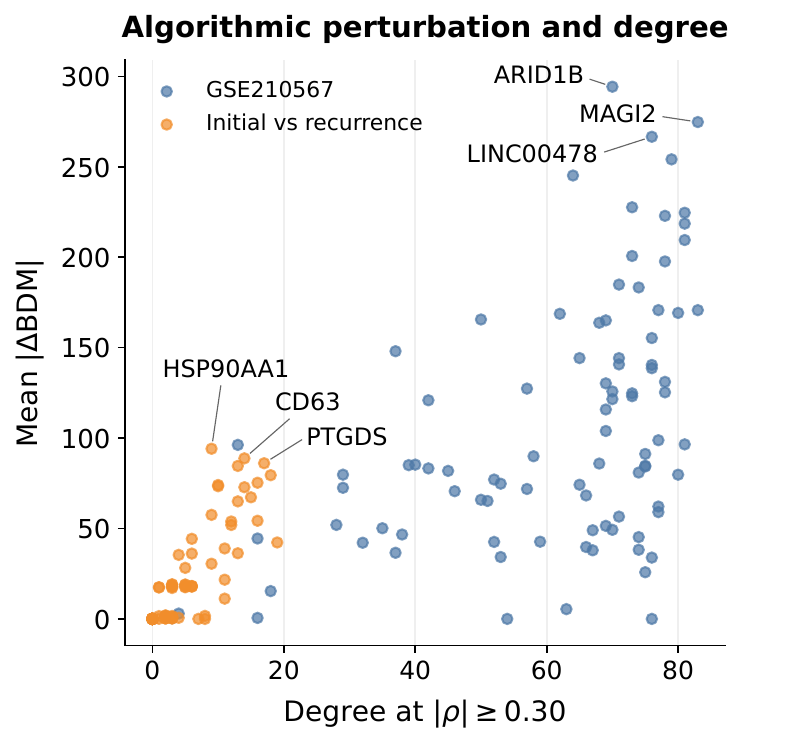}\end{subfigure}

\caption{Algorithmic perturbation signatures and network centrality. (A--B) Leading mean absolute BDM changes following node removal from union feature networks across correlation thresholds. (C) Relationship between BDM perturbation rank and conventional degree centrality. Agreement identifies genes that are both hubs and structurally disruptive. Disagreement identifies candidates whose estimated algorithmic contribution is not explained by degree alone. Sparse-network and adjacency-order limitations are addressed in the Discussion.}
\label{fig:bdm}
\end{figure}

BDM node-removal rankings for the union feature networks are shown in Figure~\ref{fig:bdm} and Table~\ref{tab:bdm}. In GSE210567, the ten largest mean absolute perturbations were \textit{ARID1B} (294.3), \textit{MAGI2} (274.7), \textit{LINC00478} (266.6), \textit{ROBO2} (254.1), \textit{LSAMP} (245.2), \textit{QKI} (227.6), \textit{ZBTB20} (224.6), \textit{NAV2} (222.9), \textit{NPAS3} (218.6), and \textit{PLCB1} (209.5). At $|\rho|\geq0.30$, the highest-degree genes were \textit{NRXN1} and \textit{MAGI2} (degree 83), followed by \textit{NPAS3, PLCB1, ZBTB20, FTX, BAI3/ADGRB3, MAML2, ROBO2}, and \textit{NAV2}. The overlap demonstrates that some BDM candidates were also hubs, while \textit{ARID1B} and \textit{LINC00478} achieved high perturbation ranks without being the two largest-degree nodes.

In initial versus recurrence, the leading BDM perturbations were \textit{HSP90AA1} (94.1), \textit{CD63} (88.8), \textit{PTGDS} (86.1), \textit{IGFBP7} (84.6), \textit{APOE} (79.4), \textit{DPP6} (75.3), \textit{MMP16} (74.0), \textit{OPCML} (73.3), \textit{FTH1} (72.8), and \textit{LHFPL3} (67.2). Degree centrality instead prioritized \textit{CLU} (19), \textit{APOE} (18), \textit{PTGDS} (17), \textit{CHI3L1} and \textit{DPP6} (16), \textit{LHFPL3} (15), and \textit{CD63, FTH1, SPARCL1}, and \textit{IGFBP7}. Thus, \textit{HSP90AA1} had the largest algorithmic perturbation but was not the largest hub, whereas \textit{CLU} was the dominant hub but not the largest BDM perturbation.

Network density varied markedly. GSE210567 contained 2,781 union edges at $|\rho|\geq0.30$, 831 at 0.50, and 22 at 0.70. Initial versus recurrence contained 207, 10, and zero edges, respectively. Consequently, high-threshold perturbation values in the initial-versus-recurrence analysis were dominated by very sparse or empty adjacency matrices. Node-level BDM shifts were weakly correlated with Shannon and compression shifts in the dense GSE network at $|\rho|=0.30$ ($\rho=0.08$--0.17), but correlations increased to $0.73$--$0.78$ at 0.50 and $0.92$--$0.96$ at 0.70. Initial-versus-recurrence node shifts were already highly correlated with simpler measures at 0.30 ($\rho=0.82$--0.91) and approached unity as the network became very sparse. Edge-level correlations were generally weak or undefined. These comparisons indicate that BDM contributed the most distinctive ranking in denser networks and that sparse-network results must be interpreted cautiously.

\begin{table}[!htbp]
\centering
\scriptsize
\caption{Algorithmic perturbation and conventional network-centrality signatures.}
\label{tab:bdm}
\begin{adjustbox}{max width=\textwidth}
\begin{tabular}{>{\raggedright\arraybackslash}p{2.7cm}>{\raggedright\arraybackslash}p{7.0cm}>{\raggedright\arraybackslash}p{6.0cm}}
\toprule
Dataset & Top BDM perturbations & Top degree centralities \\
\midrule
GSE210567 & ARID1B (294.3), MAGI2 (274.7), LINC00478 (266.6), ROBO2 (254.1), LSAMP (245.2), QKI (227.6), ZBTB20 (224.6), NAV2 (222.9), NPAS3 (218.6), PLCB1 (209.5) & MAGI2 (83), NRXN1 (83), PLCB1 (81), NPAS3 (81), FTX (81), ZBTB20 (81), MAML2 (80), BAI3 (80), ROBO2 (79), NAV2 (78) \\
Initial vs recurrence & HSP90AA1 (94.1), CD63 (88.8), PTGDS (86.1), IGFBP7 (84.6), APOE (79.4), DPP6 (75.3), MMP16 (74.0), OPCML (73.3), FTH1 (72.8), LHFPL3 (67.2) & CLU (19), APOE (18), PTGDS (17), CHI3L1 (16), DPP6 (16), LHFPL3 (15), CD63 (14), FTH1 (14), SPARCL1 (13), IGFBP7 (13) \\
\bottomrule
\end{tabular}
\end{adjustbox}
\end{table}

\subsection{Initial-to-recurrence integration prioritized disease progression-associated structural candidates}

\begin{figure}[!htbp]
\centering
\begin{subfigure}[t]{0.485\textwidth}\caption{Progression XAI signature}\includegraphics[width=\linewidth]{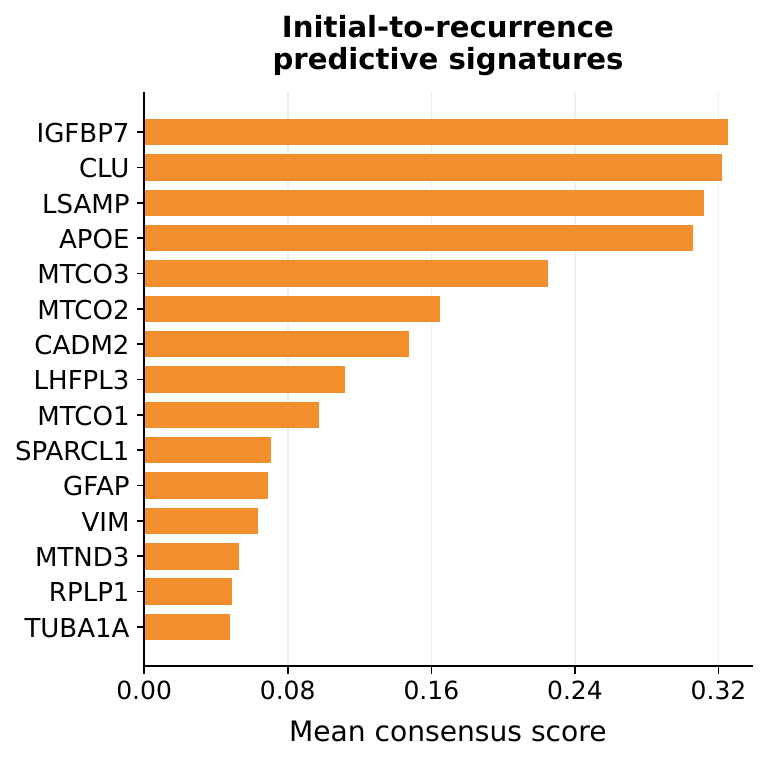}\end{subfigure}\hfill
\begin{subfigure}[t]{0.485\textwidth}\caption{BDM and centrality}\includegraphics[width=\linewidth]{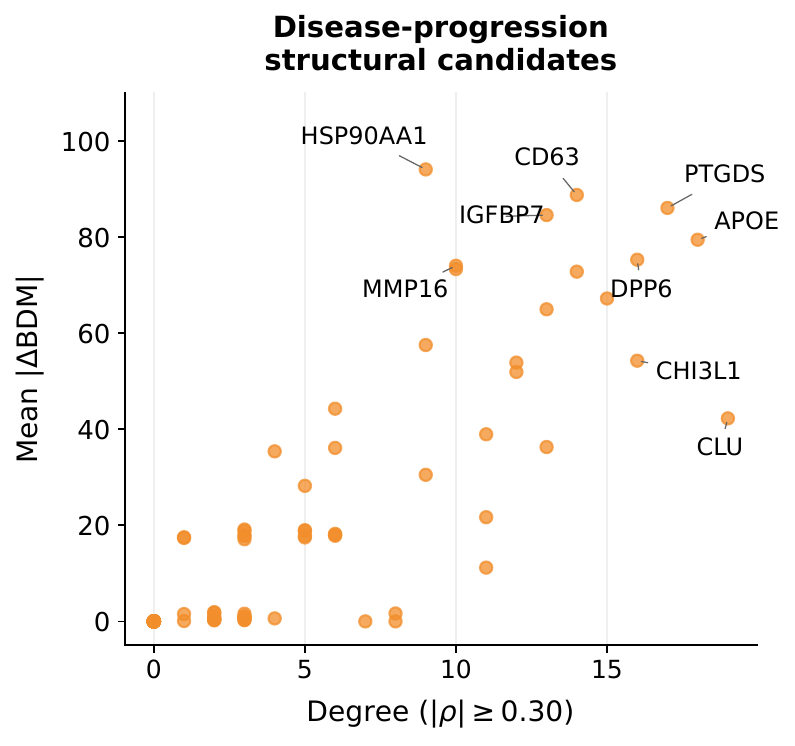}\end{subfigure}

\vspace{3mm}
\begin{subfigure}[t]{0.485\textwidth}\caption{Progression-associated network}\includegraphics[width=\linewidth]{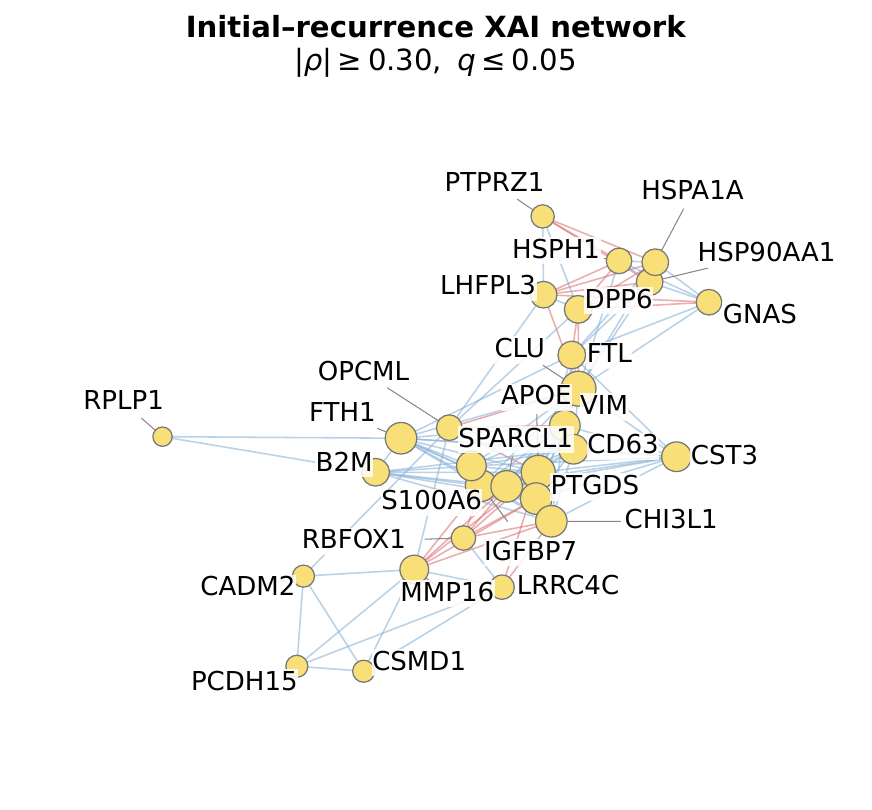}\end{subfigure}\hfill
\begin{subfigure}[t]{0.485\textwidth}\caption{Integrated progression ranking}\includegraphics[width=\linewidth]{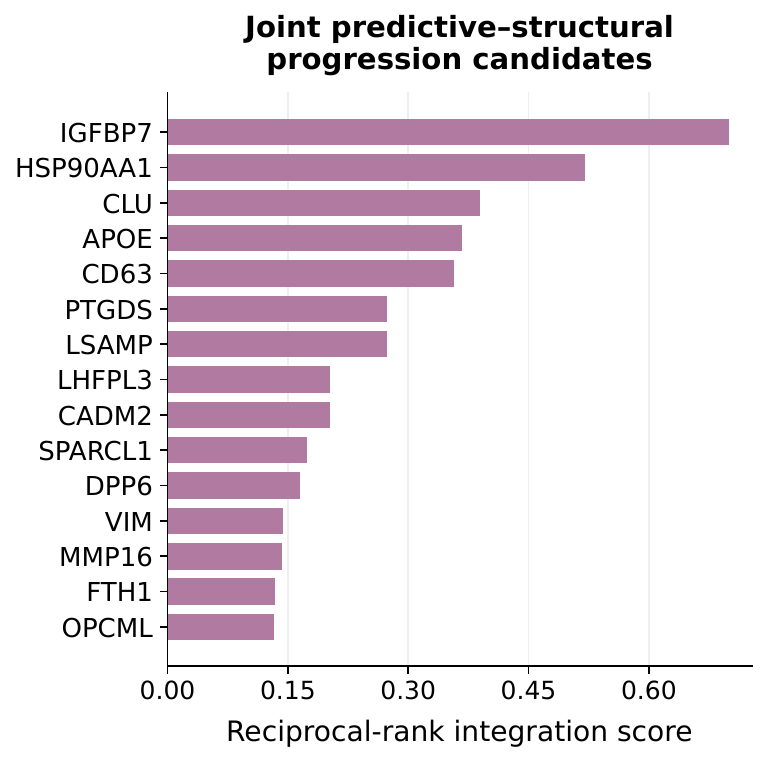}\end{subfigure}
\caption{Initial-versus-recurrence H3K27M analysis. (A) Decoder and integrated-gradient predictors. (B) Leading BDM node perturbations compared with degree. (C) Union feature association network at $|\rho|\geq0.30$ and FDR $q\leq0.05$. (D) Reciprocal-rank integration of predictive and algorithmic-structural evidence. These are progression-discriminating signatures. Absolute XAI and unsigned networks do not establish whether each gene increases or decreases at recurrence.}
\label{fig:progression}
\end{figure}

The longitudinal initial-versus-recurrence analysis is summarized in Figure~\ref{fig:progression}. Its strongest XAI union genes were \textit{IGFBP7, CLU, LSAMP, APOE, MT-CO3, MT-CO2, CADM2, LHFPL3, MT-CO1}, and \textit{SPARCL1}. Its strongest BDM candidates were \textit{HSP90AA1, CD63, PTGDS, IGFBP7, APOE, DPP6, MMP16, OPCML, FTH1}, and \textit{LHFPL3}. Reciprocal-rank integration prioritized genes that were simultaneously predictive and structurally disruptive, including \textit{IGFBP7, CLU, APOE, LSAMP, LHFPL3, DPP6, CD63, PTGDS}, and \textit{SPARCL1}. The $|\rho|\geq0.30$ graph placed \textit{CLU, APOE, PTGDS, DPP6, CHI3L1}, and \textit{LHFPL3} among the most connected progression-associated genes. This signature should be understood as an initial-versus-recurrence discriminator, not as a list of genes proven to increase at recurrence. The XAI workflow used absolute attribution, and the BDM workflow used unsigned correlation adjacency. 

\subsection{Cross-analysis integration identified recurrent neural-state and plasticity candidates}

\begin{figure}[!htbp]
\centering
\begin{subfigure}[t]{0.485\textwidth}\caption{Cross-dataset XAI ranking}\includegraphics[width=\linewidth]{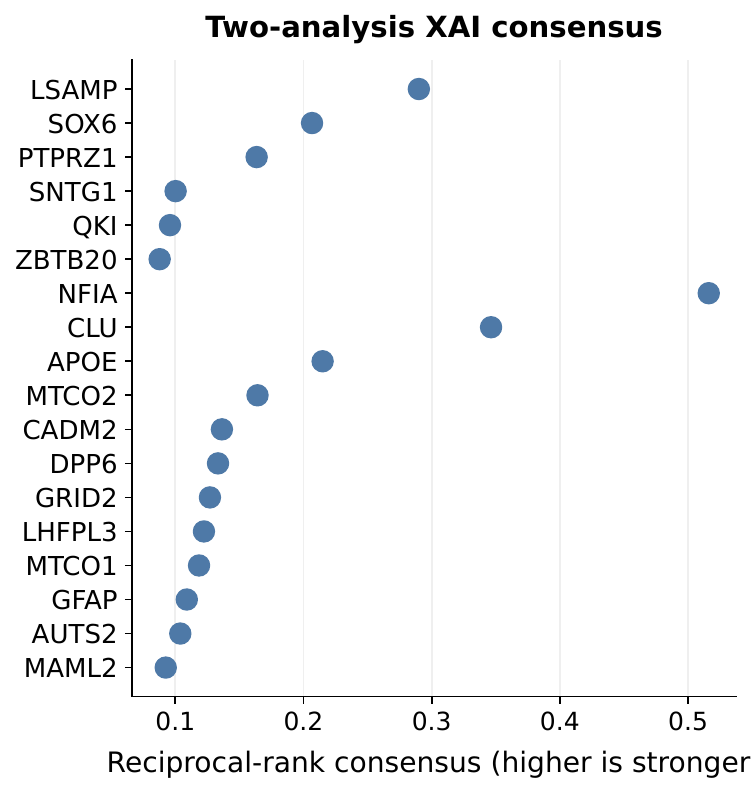}\end{subfigure}\hfill
\begin{subfigure}[t]{0.485\textwidth}\caption{XAI recurrence}\includegraphics[width=\linewidth]{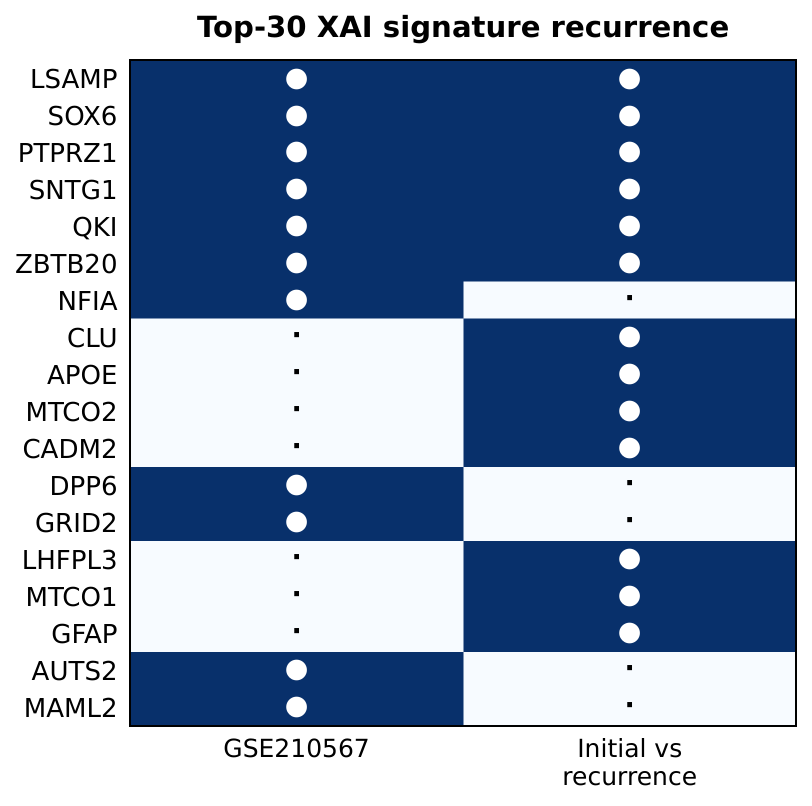}\end{subfigure}

\vspace{3mm}
\begin{subfigure}[t]{0.485\textwidth}\caption{BDM recurrence}\includegraphics[width=\linewidth]{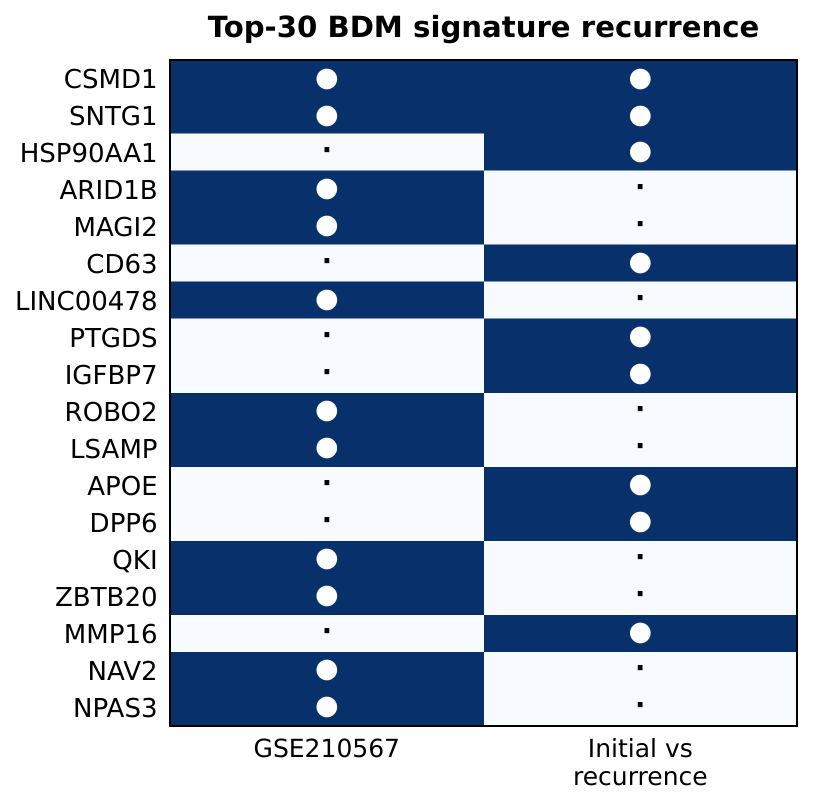}\end{subfigure}\hfill
\begin{subfigure}[t]{0.485\textwidth}\caption{Integrated XAI--BDM evidence}\includegraphics[width=\linewidth]{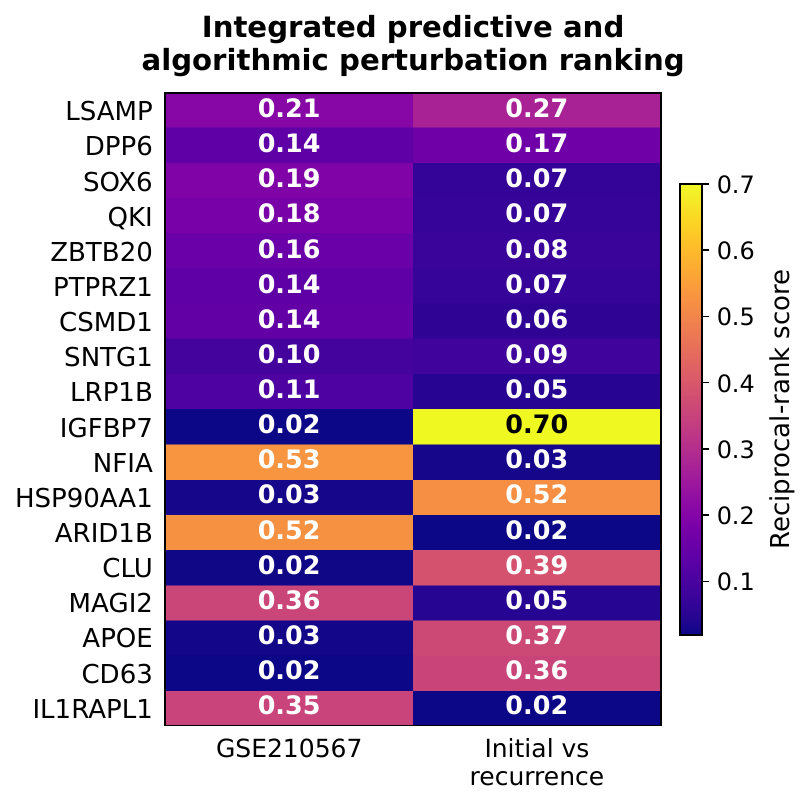}\end{subfigure}
\caption{Cross-analysis integration of predictive and structural signatures for the two retained datasets. (A) XAI consensus from per-analysis UNION ranks. (B) Presence of leading XAI candidates. (C) Presence of leading BDM perturbation candidates. (D) Dataset-specific reciprocal-rank integration. \textit{LSAMP, DPP6, SOX6, QKI, ZBTB20, PTPRZ1, CSMD1, SNTG1}, and \textit{LRP1B} met the top-rank presence criterion in both analyses.}
\label{fig:crossdataset}
\end{figure}

Cross-analysis recurrence is shown in Figure~\ref{fig:crossdataset}. Using the unchanged dataset-specific XAI and BDM ranks, \textit{LSAMP, DPP6, SOX6, QKI, ZBTB20, PTPRZ1, CSMD1, SNTG1}, and \textit{LRP1B} met the top-rank presence criterion in both retained analyses. The largest dataset-specific integrated scores remained \textit{IGFBP7} and \textit{HSP90AA1} for disease progression and \textit{NFIA} and \textit{ARID1B} for GSE210567 cell state. The cross-analysis summary therefore highlights repeated support without changing any retained per-analysis ranking.

These results support a layered interpretation. XAI recurrence identifies genes repeatedly used by the synchronization representation, while BDM recurrence identifies genes whose removal most changes estimated network description length. Their intersection yields a compact set of candidates for experimental perturbation.
\subsection{Exploratory enrichment highlighted neural and recurrence-associated functions}
\label{sec:enrichment_results}

Exploratory over-representation (using the identified candidate transition/network biomarkers-based gene enrichment analysis) identified 52 significant signature--term associations after global correction (Supplementary Figures S9--S11). In GSE210567, BDM-node candidates were enriched for axon guidance ($q=0.0143$), cell--cell adhesion ($q=0.0189$) and nervous system development ($q=0.0211$), while the XAI union was enriched for generation of neurons ($q=0.0163$). Thus, matrisome and adhesome-based neuronal program enrichment were prominent. In the initial-versus-recurrence comparison, mitochondrial RNA-processing annotations were prominent across the XAI signatures: the FASTK-family pathway reached $q=7.52\times10^{-8}$ in both latent IG and the union, driven by mitochondrial transcripts rather than FASTK-family genes. Recurrence BDM candidates additionally implicated scavenger-receptor ligand uptake ($q=0.0123$) and chaperone-mediated protein-complex assembly ($q=0.0178$). These annotation-based associations provide functional context for the computational signatures. 

\section{Discussion}

This study developed scCTM-RL-AID as a layered framework for modelling glioma cell-state decisions from single-cell transcriptomes, both static snapshots and limited, sparsely available time-series. The framework combines four analytical levels that are often studied separately: CTM synchronization-based representation learning, reinforcement-learning perturbations in the learned latent space, explainable-AI attribution of the genes supporting that representation learning, and algorithmic perturbation analysis of model-informed gene regulatory networks. Across two fully completed five-fold analyses, the principal contribution was not an improvement in classification by the RL policy. Rather, it was the construction of a replicable path from cell expression to synchronization latent state, empirical Waddington-like potential landscape, predictive feature signature, and BDM-ranked in silico intervention candidates.

This systems medicine framework places computational oncology within a complexity-informed perspective: phenotypic plasticity or cellular state-transitions reflects multi-agent interactions among developmental programs, cellular states, and their environmental constraints, rather than a fixed binary of stem cell and differentiated-like identity \citep{Filbin2018,Neftel2019,Jessa2022}. Here, an attractor denotes a persistent pattern approached by an evolving system in state space. A dense region of a static embedding is only an attractor-like candidate, not evidence that its cells dynamically converge or remain trapped there. This distinction allows the cybernetics of tumour ecosystems, differentiation continua, and maladaptive state persistence to motivate testable candidate transition genes and hypotheses without claiming that the present analysis demonstrates specific dynamical features such as criticality, chaos, or a phase transition. Connecting representation learning and network perturbation analysis can help prioritize such candidates and in silico predictions, but prospective experimental, temporal and intervention data are required to turn static state discrimination into cell-fate forecasting or recurrence early warning signals.

\subsection{CTM synchronization provided a high-capacity representation, whereas RL acted mainly as a perturbation probe}

The CTM autoencoder separated GSE210567 stem-like and differentiated cells and initial versus recurrent H3K27M cells with near-ceiling AUROC. These two tasks probe complementary ecology: cell-state discrimination within H3K27M malignant cells and longitudinal disease-stage discrimination. Glioma studies have repeatedly shown that developmental and malignant programs are distributed across plastic states rather than forming perfectly discrete categories \citep{Filbin2018,Neftel2019,Jessa2022,Sussman2026}. The strong retained results therefore support predictive representation learning while still requiring patient-held-out replication in future studies, should a larger patient cohort be available.

The deterministic RL rollout preserved classification but did not improve it materially and increased reconstruction error in every dataset. This result constrains the interpretation of the policy. The present reward was optimized toward an NMF-derived within-class prototype while retaining the correct class and limiting action magnitude. Such a policy can reveal directions in synchronization space that are computationally compatible with a target state, but it is not automatically expected to improve a classifier already operating near its performance ceiling. The increased reconstruction error indicates that the learned action displaced cells away from exact expression reconstruction as it optimized the prototype-directed reward. Accordingly, the RL component is best described as a \emph{latent intervention probe}: it tests whether action-penalized movements can be learned in the CTM state space and provides perturbed latent coordinates for post hoc landscape analysis. It should not be described as an experimentally validated cell-fate controller.

CTM differs conceptually from a conventional Transformer because its synchronization representation is formed through neuron-level histories across an internal thought axis \citep{Darlow2025}. This offers an attractive mechanism for attractor landscape reconstruction, representing repeated internal computation from a static transcriptomic input into a complex dynamics forecasting toolkit when combined with complex networks. However, the present study did not perform a matched Transformer or world model/jepa benchmark, measure asymptotic scaling, or compare attention maps with synchronization matrices. Multi-omics integration such as data from chromatin accessibility will be a necessary future step. The results therefore support feasibility and interpretability of CTM synchronization in single-cell modelling, not general superiority or greater scalability over Transformers, recurrent neural networks, variational models, physics-informed neural differential equations, or optimal-transport approaches.

\subsection{Empirical potential landscapes captured class-conditioned latent organization, not measured biological velocity}

The out-of-fold latent embeddings generated Waddington-like density landscapes in both retained analyses. GSE210567 displayed the clearest basin separation, with PHATE-derived potential distinguishing the two classes with AUROC 0.91. Initial versus recurrence was almost perfectly separable by the full CTM output but only moderately separated in PHATE potential. This discordance is important. A nonlinear classifier can distribute predictive information across many latent dimensions even when a two- or three-dimensional visualization does not yield compact class clusters.

The potential $U=-\log p$ provides an empirical energy-like representation of latent occupancy: dense regions are assigned lower potential and sparse regions higher potential. This construction is consistent with the qualitative Waddington metaphor and with the dynamical-systems interpretation of phenotypes as attractor-like regions \citep{Waddington1957,Huang2009}. It does not, by itself, estimate a nonequilibrium biological landscape, transition probability, lineage genealogy, or vector field generated by gene-regulatory kinetics. The plotted field $-\nabla U$ is the gradient of a kernel-density surface, not RNA velocity. Likewise, the potential-ranked coordinate is not pseudotime because it lacks a lineage root and transition operator. Further, there may be a more complex plastic spectrum of hybrid and transient cellular fates or subtypes, including hybrid lineage progenitors and myeloid subtypes skewing conventional fate bifurcations. These distinctions are central to a defensible interpretation of attractor landscape reconstruction from static data.

A future extension could combine synchronization latent states with longitudinal sampling, metabolic labelling, epigenomics, clonal lineage tracing,  chromatin accessibility, RNA velocity, or optimal transport. Those data would allow the density-derived potential to be compared with an independently estimated transition field and could support a landscape-flux decomposition rather than a purely gradient field. The existing complex systems framework is nevertheless useful as a descriptive mapping of Waddington-like landscapes: it identifies where predictive states occupy the learned manifold and provides a coordinate system in which computational perturbations can be visualized.

\subsection{XAI identified complementary reconstruction, synchronization, and cross-analysis signatures}

Decoder $L_2$, reconstruction IG, and latent IG interrogated different parts of the model. Decoder norms measured how strongly latent decoder units contributed to reconstruction of each gene. Reconstruction IG quantified the sensitivity of reconstruction fidelity to each input feature, whereas latent IG quantified the contribution of each feature to synchronization-state magnitude. The partial convergence of these rankings is more informative than any one ranking alone: genes recurrent across methods and folds were used both to form the learned latent representation and to reconstruct the observed transcriptome.

In GSE210567, the XAI union emphasized developmental and neural-state genes including \textit{NFIA, SOX6, ERBB4, PTPRZ1, DPP6, GRID2}, and \textit{AUTS2}, together with adhesion and signalling genes such as \textit{IL1RAPL1} and \textit{ADGRB3/BAI3}. This signature is consistent with the idea that malignant glioma states reuse neural developmental, adhesion (adhesome-matrisome), cytoskeletal, and membrane-signalling programs, although the present absolute-attribution analysis does not determine whether a gene is activated or repressed in a particular cell state.

The initial-versus-recurrence comparison highlighted \textit{IGFBP7, CLU, APOE, SPARCL1, PTGDS, CD63}, mitochondrial transcripts, and neural-identity genes. This combination suggests that recurrence prediction used extracellular-matrix and glial-support programs, cellular stress and proteostasis, metabolic state, and residual neural lineage identity. The result should not be converted directly into a recurrence-upregulated gene list, because absolute IG and unsigned decoder scores lack direction. It is instead a predictive progression signature: these features carried information that the trained model used to distinguish initial from recurrent cells. Patient-level differential expression and effect directions should be added before biological claims about recurrence-associated increases or decreases are made.

Several identified genes have independent relevance to glioma state biology. \textit{PTPRZ1}, which recurred within the top XAI ranks of both retained analyses, participates in glioma stem-cell signalling, invasion, and tumour maintenance \citep{Shi2017,Fujikawa2017,Bhaduri2020}. This external evidence increases the biological plausibility of the ranking.

Across the retained analyses, \textit{LSAMP, SOX6, PTPRZ1, SNTG1, QKI}, and \textit{ZBTB20} appeared within the top-30 XAI UNION rankings of both datasets. Cross-analysis recurrence is useful because a gene can be important in one dataset because of platform, sampling, or label-specific structure. Repetition across cell-state and disease-stage analyses makes a candidate more attractive for follow-up, while preserving dataset-specific candidates such as \textit{IGFBP7} in progression and \textit{NFIA} in GSE210567.

\subsection{BDM perturbation analysis complemented, but did not replace, centrality}

The BDM analysis asked a different question from XAI. XAI measured model sensitivity to expression features. BDM measured the change in estimated algorithmic description length when a node or edge was removed from a thresholded, model-informed association network. The largest GSE210567 perturbations included \textit{ARID1B, MAGI2, LINC00478, ROBO2, LSAMP, QKI, ZBTB20, NAV2, NPAS3}, and \textit{PLCB1}. Initial-versus-recurrence perturbations prioritized \textit{HSP90AA1, CD63, PTGDS, IGFBP7, APOE, DPP6, MMP16, OPCML, FTH1}, and \textit{LHFPL3}.

Comparison with degree centrality showed both convergence and divergence. Highly connected genes such as \textit{MAGI2}, \textit{DPP6}, and \textit{LHFPL3} also produced large BDM shifts, whereas \textit{ARID1B}, \textit{LINC00478}, and \textit{HSP90AA1} achieved high algorithmic perturbation ranks without being the largest hubs. This is the intended contribution of algorithmic perturbation analysis: it can prioritize changes in global estimated description length that are not reducible to the number of incident edges \citep{Zenil2019,Zenil2018BDM,Uthamacumaran2022}. Nevertheless, the present results also expose important constraints. BDM correlated strongly with Shannon and compression perturbations as networks became sparse, and the high-threshold initial-versus-recurrence network contained few or no edges. Under those conditions, many measures converge on the same dominant property---edge removal from a small binary object.

BDM edge perturbation analysis identified a distinct set of high-impact interactions across the two glioma settings (Supplementary Figs.~S7--S8). In GSE210567, the strongest perturbation-sensitive edges included \textit{LINC00478--NTM}, \textit{PDE4B--MAGI2}, \textit{MALAT1--ARID1B}, \textit{AUTS2--MAGI2}, \textit{FTX--ARID1B}, \textit{LPHN3--CTNND2}, \textit{NCAM2--CTNND2}, \textit{AUTS2--ARID1B}, and \textit{PTPRZ1--CTNND2}. In the initial-versus-recurrence network, leading signatures included \textit{CLU--HSP90AA1}, \textit{CD63--NDUFA4}, \textit{CST3--CD63}, \textit{IGFBP7--FTH1}, \textit{APOE--MMP16}, \textit{DPP6--HSPH1}, \textit{CADM2--DSCAM}, \textit{CADM2--CSMD1}, and \textit{DPP6--AUTS2}. Notably, several of these interactions converge on chromatin/epigenetic regulation and neuronal or developmental-state programs, with recurrent involvement of \textit{ARID1B}, \textit{AUTS2}, \textit{CTNND2}, \textit{DPP6}, and \textit{CADM2}, suggesting candidate perturbation-sensitive network constraints associated with cell-state plasticity and recurrence (See Supplementary Information). Again, we stress that these are candidate perturbation-sensitive network constraints/interactions,because BDM perturbation identifies structurally influential edges but does not by itself prove causal biological regulation without future experimental validation.

Algorithmic information dynamics (AID) supplies a principled causal calculus in silico for generative description and compression, but exact Kolmogorov complexity is semi-computable. BDM is therefore an algorithmic complexity estimator when combined with perturbation analysis, built from small-block algorithmic-probability approximations and decomposition, while GZIP and LZW quantify compressor-dependent regularities and infer generative structures (programs) steering the complex system\citep{Zenil2018BDM,Zenil2019,Uthamacumaran2022}. Combining these quantities with learned representations is useful for comparing structural sensitivity, provided estimator dependence, encoding choices, and the distinction between computational and biological intervention remain explicit.

BDM of an adjacency matrix can also depend on node ordering because block decomposition operates on the matrix representation. The fixed gene order used here makes within-analysis perturbations comparable but does not establish permutation invariance. Future work should report perturbations averaged across biologically constrained and random node permutations, compare adjacency BDM with canonical graph encodings, and use degree-preserving null networks. Threshold-free weighted encodings or multilayer networks could reduce the abrupt information loss observed at $|\rho|=0.50$ and 0.70.

For these reasons, BDM is best treated as a complementary prioritization layer rather than a standalone causal test. In the present study, ``causal'' denotes the formal intervention logic of removing an element and quantifying the resulting change in an algorithmic-information estimator. Biological causality requires that perturbing the candidate changes cell state or transition propensity in an appropriate experimental system. The strongest candidates for such experiments are therefore genes supported jointly by XAI recurrence, BDM perturbation, conventional network position, and external systems ecology.

\subsection{Integrated signatures nominate progression and plasticity candidates}

Reciprocal-rank integration produced a compact hierarchy of candidates across the two retained analyses. \textit{LSAMP, DPP6, SOX6, QKI, ZBTB20, PTPRZ1, CSMD1, SNTG1}, and \textit{LRP1B} met the top-rank presence criterion in both analyses. These genes span neural excitability and membrane organization, receptor and adhesion signalling, developmental transcriptional regulation, and state-maintenance programs. Their recurrence supports a view of glioma plasticity and cell fate transitions as distributed across interacting network modules rather than governed by a single marker or network orchestration layer.

For disease progression, the combined evidence placed \textit{IGFBP7, CLU, APOE, LHFPL3, DPP6, CD63, PTGDS}, and \textit{SPARCL1} among the most relevant predictive and structural candidates. \textit{HSP90AA1} was the largest BDM perturbation, whereas \textit{CLU} was the largest degree hub and the dominant decoder/reconstruction feature. This separation illustrates why the integrated framework is useful: a predictive hub, an algorithmically disruptive element, and a recurrent cross-dataset feature need not be the same gene. These complementary roles can guide a staged validation strategy, for example prioritizing one set for biomarker evaluation and another for perturbational screening.

\subsection{Limitations}

Several limitations define the present evidential boundary. First, cross-validation was stratified at the cell level, not grouped by patient. Cells from the same tumour may therefore appear in training and test partitions, allowing patient-specific expression structure to inflate generalization estimates. The near-perfect initial-versus-recurrence performance must be replicated under patient-held-out validation. Second, the study did not include matched Transformer, recurrent, variational, neural-ODE, optimal-transport, RNA-velocity, or conventional trajectory-inference baselines.

Third, the landscape is density-derived and does not recover measured temporal direction, a physical energy, or nonequilibrium flux. Future experiments can investigate these alternative physics-driven models using  neural-ode/pdes and PINNs, as well as alternative network inference methods such as attractor inference from Kauffman's random boolean networks. Fourth, XAI used an all-zero baseline, and absolute attributions. Alternative biologically matched baselines, signed class-conditional attribution, and attribution-stability analysis are needed. Fifth, the inferred networks are undirected Spearman association networks, not directed gene-regulatory networks, and can reflect co-expression, cell composition, genomic location, and technical covariation. Sixth, BDM perturbation values depend on graph encoding, node order, binary threshold, and network size, and the strongest comparator correlations occurred in sparse networks. Further, the integrated reciprocal-rank score was a transparent post hoc prioritization rule rather than a prespecified learned objective.

Finally, no in silico candidate was perturbed experimentally, and it must be stressed again that this is a computational and systems oncology study only. CRISPR precision gene editing/interference or activation, drug perturbation, lineage-resolved organoid models, and longitudinal patient-derived systems will be required to determine whether high-ranked genes alter attractor occupancy, state-transition rate/probability, recurrence-associated state, or treatment response. Until then, the framework supports reproducible in silico hypothesis generation rather than proof of causal control or generative programs of glioma ecosystems.

\subsection{Future directions}

The immediate computational priorities are patient-grouped nested cross-validation, stabilized full-precision RL, direct architecture and trajectory baselines, signed class-specific XAI, and permutation-robust network complexity. A second generation of the model could incorporate BDM or network-structure change into a constrained policy objective, while maintaining a separate held-out evaluation to avoid circular validation. More importantly, transition-supervised data could replace static prototype rewards. Learning from paired perturbation-response profiles, longitudinal samples, clonal descendants, or optimal-transport couplings would allow the actor to model observed state changes rather than proximity to a class centroid.

A complementary research direction is to compare the CTM representation with masked joint-embedding predictive models, neural ODEs or latent SDEs, reservoir computing, and Transformers under the same patient-grouped evaluation. With longitudinal or perturbation-resolved observations, evolving gene--cell graphs could connect latent-state changes to network topology, entropy, compression, and algorithmic perturbation sensitivity across scales. Early-warning indicators such as increasing variance or lagged dependence would then need validation against observed transitions and appropriate null models. Such extensions could eventually inform differentiation or reprogramming experiments, but the present transcriptomic rankings do not establish effective gene-editing interventions, preventive vaccines, or clinically validated recurrence-risk signals.

Experimental follow-up should test a compact panel representing complementary evidence classes: cross-analysis integrated candidates such as \textit{LSAMP, DPP6, PTPRZ1}, and \textit{SOX6}. Disease progression-associated predictive candidates such as \textit{IGFBP7, CLU, APOE}, and \textit{SPARCL1}, and high-BDM candidates such as \textit{ARID1B, HSP90AA1, MAGI2}, and \textit{CD63}. Measuring whether these perturbations change cell-state composition, latent basin occupancy, differentiation propensity, and treatment sensitivity would directly test the central hypothesis of scCTM-RL-AID.

\section{Conclusion}

scCTM-RL-AID connects synchronization-based neural representation and manifold learning with empirical latent landscapes, explainable feature attribution, and algorithmic perturbation analysis. In two fully completed glioma single-cell analyses, CTM-AE provided strong state and disease-stage prediction, whereas the RL policy preserved classification but did not improve reconstruction. The RL component is therefore most defensibly interpreted as a computational latent perturbation probe. Decoder $L_2$, reconstruction IG, and latent IG recovered reproducible predictive signatures, while BDM and centrality provided complementary rankings of network-structural importance. Cross-analysis integration recurrently highlighted \textit{LSAMP, DPP6, SOX6, QKI, ZBTB20, PTPRZ1, CSMD1, SNTG1}, and \textit{LRP1B}, and the progression analysis identified a set containing \textit{IGFBP7, CLU, APOE, LHFPL3, DPP6, CD63, PTGDS}, and \textit{SPARCL1}. Edge-level BDM further highlighted perturbation-sensitive interactions involving \textit{ARID1B, AUTS2, CTNND2, NRXN1, DPP6}, and \textit{CADM2}, complementing the prominent progression-associated \textit{NRXN3} signal. These node- and edge-level signatures nominate neural-developmental, adhesion, and chromatin-associated network interactions as candidate regulators of glioma cell-state plasticity. These candidates define experimentally testable hypotheses about glioma plasticity and recurrence. The framework does not yet establish biological trajectories or causality, but it provides an explicit and reproducible bridge from predictive modelling to latent-state organization and in silico intervention prioritization.

\section*{Data and code availability}
The original code and algorithms are available at the project GitHub repository: \url{https://github.com/Abicumaran/scCTM}. The repository will be made publicly available upon publication of the paper. The official Continuous Thought Machine implementation is cited in the bibliography \citep{SakanaCTMCode}. GSE210567 contains single-cell RNA-sequencing data from the Jessa H3K27M midline-glioma atlas. The present malignant-cell stem-like-versus-differentiated analysis retained 18,490 cells \citep{Jessa2022}. The initial-versus-recurrence analysis used a capped subset of stratified 20,000 H3K27M patient cells from the Sussman longitudinal pediatric high-grade-glioma single-nucleus atlas \citep{Sussman2026}. Its source data are available through HTAN/dbGaP (phs002371), with processed data in CELLxGENE collection \url{https://cellxgene.cziscience.com/collections/9ceda3d2-a55e-4b82-a91a-0190d947459a}. These counts describe the analysed subsets, not the total source atlases.

\section*{Competing interests}
The authors declare no competing interests.

\input{references_frozen.tex}

\clearpage

\end{document}

%% file: references_frozen.tex

%% file: main__11_.bbl
\begin{thebibliography}{31}
\providecommand{\natexlab}[1]{#1}
\providecommand{\url}[1]{\texttt{#1}}
\expandafter\ifx\csname urlstyle\endcsname\relax
  \providecommand{\doi}[1]{doi: #1}\else
  \providecommand{\doi}{doi: \begingroup \urlstyle{rm}\Url}\fi

\bibitem[Benjamini and Hochberg(1995)]{Benjamini1995}
Yoav Benjamini and Yosef Hochberg.
\newblock Controlling the false discovery rate: a practical and powerful
  approach to multiple testing.
\newblock \emph{Journal of the Royal Statistical Society Series B}, 57\penalty0
  (1):\penalty0 289--300, 1995.

\bibitem[Bhaduri et~al.(2020)Bhaduri, Di~Lullo, Jung, M{\"u}ller, Crouch,
  Espinosa, Ozawa, Alvarado, Spatazza, Cadwell, et~al.]{Bhaduri2020}
Aparna Bhaduri, Elizabeth Di~Lullo, Donglian Jung, S{\"o}ren M{\"u}ller,
  Elizabeth~E. Crouch, Cristina~S. Espinosa, Tomoko Ozawa, Beatriz Alvarado,
  Julien Spatazza, Cathryn~R. Cadwell, et~al.
\newblock Outer radial glia-like cancer stem cells contribute to heterogeneity
  of glioblastoma.
\newblock \emph{Cell Stem Cell}, 26:\penalty0 48--63.e6, 2020.
\newblock \doi{10.1016/j.stem.2019.11.015}.

\bibitem[Darlow et~al.(2025)Darlow, Regan, Risi, Seely, and Jones]{Darlow2025}
Luke Darlow, Ciaran Regan, Sebastian Risi, Jeffrey Seely, and Llion Jones.
\newblock Continuous thought machines.
\newblock \emph{arXiv preprint arXiv:2505.05522}, 2025.
\newblock \doi{10.48550/arXiv.2505.05522}.

\bibitem[Filbin et~al.(2018)Filbin, Tirosh, Hovestadt, Shaw, Escalante,
  Mathewson, Neftel, Frank, Pelton, Hebert, et~al.]{Filbin2018}
Mariella~G. Filbin, Itay Tirosh, Volker Hovestadt, Mara~L. Shaw, Leila~E.
  Escalante, Nathan~D. Mathewson, Cyril Neftel, Noah Frank, Kristine Pelton,
  Christine~M. Hebert, et~al.
\newblock Developmental and oncogenic programs in h3k27m gliomas dissected by
  single-cell rna-seq.
\newblock \emph{Science}, 360:\penalty0 331--335, 2018.
\newblock \doi{10.1126/science.aao4750}.

\bibitem[Fujikawa et~al.(2017)Fujikawa, Sugawara, Tanaka, et~al.]{Fujikawa2017}
Akihiro Fujikawa, Hiroshi Sugawara, Tomoko Tanaka, et~al.
\newblock Targeting ptprz inhibits stem cell-like properties and tumorigenicity
  in glioblastoma cells.
\newblock \emph{Scientific Reports}, 7:\penalty0 5609, 2017.
\newblock \doi{10.1038/s41598-017-05931-8}.

\bibitem[Guetta-Terrier et~al.(2023)Guetta-Terrier, Karambizi, Akosman,
  et~al.]{GuettaTerrier2023}
Charlotte Guetta-Terrier, David Karambizi, Bedia Akosman, et~al.
\newblock Chi3l1 is a modulator of glioma stem cell states and a therapeutic
  target in glioblastoma.
\newblock \emph{Cancer Research}, 83\penalty0 (12):\penalty0 1984--1999, 2023.
\newblock \doi{10.1158/0008-5472.CAN-21-3629}.

\bibitem[Haghverdi et~al.(2016)Haghverdi, B{\"u}ttner, Wolf, Buettner, and
  Theis]{Haghverdi2016}
Laleh Haghverdi, Maren B{\"u}ttner, F.~Alexander Wolf, Florian Buettner, and
  Fabian~J. Theis.
\newblock Diffusion pseudotime robustly reconstructs lineage branching.
\newblock \emph{Nature Methods}, 13:\penalty0 845--848, 2016.
\newblock \doi{10.1038/nmeth.3971}.

\bibitem[Huang et~al.(2005)Huang, Eichler, Bar-Yam, and Ingber]{Huang2005}
Sui Huang, Gabriel Eichler, Yaneer Bar-Yam, and Donald~E. Ingber.
\newblock Cell fates as high-dimensional attractor states of a complex gene regulatory network.
\newblock \emph{Physical Review Letters}, 94(12):128701, 2005.
\newblock \doi{10.1103/PhysRevLett.94.128701}.

\bibitem[Huang et~al.(2009)Huang, Ernberg, and Kauffman]{Huang2009}
Sui Huang, Ingemar Ernberg, and Stuart Kauffman.
\newblock Cancer attractors: a systems view of tumors from a gene network
  dynamics and developmental perspective.
\newblock \emph{Seminars in Cell and Developmental Biology}, 20\penalty0
  (7):\penalty0 869--876, 2009.
\newblock \doi{10.1016/j.semcdb.2009.07.003}.

\bibitem[Jessa et~al.(2022)Jessa, Mohammadnia, Harutyunyan, et~al.]{Jessa2022}
Selin Jessa, Abdulshakour Mohammadnia, Ashot~S. Harutyunyan, et~al.
\newblock K27m in canonical and noncanonical h3 variants occurs in distinct
  oligodendroglial cell lineages in brain midline gliomas.
\newblock \emph{Nature Genetics}, 54:\penalty0 1865--1880, 2022.
\newblock \doi{10.1038/s41588-022-01205-w}.

\bibitem[Kauffman(1969)]{Kauffman1969}
Stuart~A. Kauffman.
\newblock Metabolic stability and epigenesis in randomly constructed genetic nets.
\newblock \emph{Journal of Theoretical Biology}, 22(3):437--467, 1969.
\newblock \doi{10.1016/0022-5193(69)90015-0}.

\bibitem[Kuleshov et~al.(2016)]{Kuleshov2016}
Maxim~V. Kuleshov, Matthew~R. Jones, Andrew~D. Rouillard, et~al.
\newblock Enrichr: a comprehensive gene set enrichment analysis web server 2016 update.
\newblock \emph{Nucleic Acids Research}, 44(W1):W90--W97, 2016.
\newblock \doi{10.1093/nar/gkw377}.

\bibitem[Lee and Seung(1999)]{LeeSeung1999}
Daniel~D. Lee and H.~Sebastian Seung.
\newblock Learning the parts of objects by non-negative matrix factorization.
\newblock \emph{Nature}, 401:\penalty0 788--791, 1999.
\newblock \doi{10.1038/44565}.

\bibitem[McInnes et~al.(2018)McInnes, Healy, and Melville]{McInnes2018}
Leland McInnes, John Healy, and James Melville.
\newblock Umap: Uniform manifold approximation and projection for dimension
  reduction.
\newblock \emph{arXiv preprint arXiv:1802.03426}, 2018.
\newblock \doi{10.48550/arXiv.1802.03426}.

\bibitem[Moon et~al.(2019)Moon, van Dijk, Wang, Gigante, Burkhardt, Chen, Yim,
  Elzen, Hirn, Coifman, Ivanova, Wolf, and Krishnaswamy]{Moon2019}
Kevin~R. Moon, David van Dijk, Zheng Wang, Scott Gigante, Daniel~B. Burkhardt,
  William~S. Chen, Kristina Yim, Antonia van~den Elzen, Matthew~J. Hirn,
  Ronald~R. Coifman, Natalia~B. Ivanova, Guy Wolf, and Smita Krishnaswamy.
\newblock Visualizing structure and transitions in high-dimensional biological
  data.
\newblock \emph{Nature Biotechnology}, 37:\penalty0 1482--1492, 2019.
\newblock \doi{10.1038/s41587-019-0336-3}.

\bibitem[Neftel et~al.(2019)Neftel, Laffy, Filbin, Hara, Shore, Rahme, Richman,
  Silverbush, Shaw, Hebert, et~al.]{Neftel2019}
Cyril Neftel, John Laffy, Mariella~G. Filbin, Takahide Hara, Marni~E. Shore,
  Gilbert~J. Rahme, Alexander~R. Richman, Dana Silverbush, Mara~L. Shaw,
  Christine~M. Hebert, et~al.
\newblock An integrative model of cellular states, plasticity, and genetics for
  glioblastoma.
\newblock \emph{Cell}, 178:\penalty0 835--849.e21, 2019.
\newblock \doi{10.1016/j.cell.2019.06.024}.

\bibitem[{Sakana AI}(2025)]{SakanaCTMCode}
{Sakana AI}.
\newblock Continuous thought machines: official implementation.
\newblock GitHub repository, 2025.
\newblock URL \url{https://github.com/SakanaAI/continuous-thought-machines}.
\newblock Accessed July 2026.

\bibitem[Seal et~al.(2023)]{Seal2023}
Ruth~L. Seal, Bryony Braschi, Kristian Gray, Tamsin~E.~M. Jones, Susan Tweedie, Liora Haim-Vilmovsky, and Elspeth~A. Bruford.
\newblock Genenames.org: the HGNC resources in 2023.
\newblock \emph{Nucleic Acids Research}, 51(D1):D1003--D1009, 2023.
\newblock \doi{10.1093/nar/gkac888}.

\bibitem[Shi et~al.(2017)Shi, Ping, Zhou, He, Chen, and Bian]{Shi2017}
Yi~Shi, Yifan~F. Ping, Wei Zhou, Zhi-Cai He, Chun Chen, and Xiu-Wu Bian.
\newblock Tumour-associated macrophages secrete pleiotrophin to promote ptprz1
  signalling in glioblastoma stem cells for tumour growth.
\newblock \emph{Nature Communications}, 8:\penalty0 15080, 2017.
\newblock \doi{10.1038/ncomms15080}.

\bibitem[Soler-Toscano et~al.(2014)Soler-Toscano, Zenil, Delahaye, and
  Gauvrit]{SolerToscano2014}
Fernando Soler-Toscano, Hector Zenil, Jean-Paul Delahaye, and Nicolas Gauvrit.
\newblock Calculating kolmogorov complexity from the output frequency
  distributions of small turing machines.
\newblock \emph{PLoS ONE}, 9\penalty0 (5):\penalty0 e96223, 2014.
\newblock \doi{10.1371/journal.pone.0096223}.

\bibitem[Sundararajan et~al.(2017)Sundararajan, Taly, and
  Yan]{Sundararajan2017}
Mukund Sundararajan, Ankur Taly, and Qiqi Yan.
\newblock Axiomatic attribution for deep networks.
\newblock In \emph{Proceedings of the 34th International Conference on Machine
  Learning}, volume~70, pages 3319--3328, 2017.

\bibitem[Sussman et~al.(2026)]{Sussman2026}
Jonathan~H. Sussman et~al.
\newblock A longitudinal single-cell and spatial multiomic atlas of pediatric
  high-grade glioma.
\newblock \emph{Cell Reports Medicine}, 7\penalty0 (5):\penalty0 102766, 2026.
\newblock \doi{10.1016/j.xcrm.2026.102766}.

\bibitem[Sutton et~al.(1999)Sutton, McAllester, Singh, and Mansour]{Sutton1999}
Richard~S. Sutton, David McAllester, Satinder Singh, and Yishay Mansour.
\newblock Policy gradient methods for reinforcement learning with function
  approximation.
\newblock \emph{Advances in Neural Information Processing Systems}, 12, 1999.

\bibitem[Traag et~al.(2019)Traag, Waltman, and van Eck]{Traag2019}
V.~A. Traag, L.~Waltman, and N.~J. van Eck.
\newblock From louvain to leiden: guaranteeing well-connected communities.
\newblock \emph{Scientific Reports}, 9:\penalty0 5233, 2019.
\newblock \doi{10.1038/s41598-019-41695-z}.

\bibitem[Trapnell et~al.(2014)Trapnell, Cacchiarelli, Grimsby, Pokharel, Li,
  Morse, Lennon, Livak, Mikkelsen, and Rinn]{Trapnell2014}
Cole Trapnell, Davide Cacchiarelli, Jonna Grimsby, Prapti Pokharel, Shuqiang
  Li, Michael Morse, Niall~J. Lennon, Kenneth~J. Livak, Tarjei~S. Mikkelsen,
  and John~L. Rinn.
\newblock The dynamics and regulators of cell fate decisions are revealed by
  pseudotemporal ordering of single cells.
\newblock \emph{Nature Biotechnology}, 32:\penalty0 381--386, 2014.
\newblock \doi{10.1038/nbt.2859}.

\bibitem[Uthamacumaran(2021)]{Uthamacumaran2021}
Abicumaran Uthamacumaran.
\newblock A review of dynamical systems approaches for the detection of chaotic attractors in cancer networks.
\newblock \emph{Patterns}, 2(4):100226, 2021.
\newblock \doi{10.1016/j.patter.2021.100226}.

\bibitem[Uthamacumaran and Craig(2022)]{Uthamacumaran2022}
Abicumaran Uthamacumaran and Morgan Craig.
\newblock Algorithmic reconstruction of glioblastoma network complexity.
\newblock \emph{iScience}, 25\penalty0 (5):\penalty0 104179, 2022.
\newblock \doi{10.1016/j.isci.2022.104179}.

\bibitem[Waddington(1957)]{Waddington1957}
Conrad~H. Waddington.
\newblock \emph{The Strategy of the Genes: A Discussion of Some Aspects of
  Theoretical Biology}.
\newblock George Allen and Unwin, London, 1957.

\bibitem[Wang et~al.(2011)Wang, Zhang, Xu, and Wang]{Wang2011}
Jin Wang, Kun Zhang, Li Xu, and Erkang Wang.
\newblock Quantifying the Waddington landscape and biological paths for development and differentiation.
\newblock \emph{Proceedings of the National Academy of Sciences}, 108(20):8257--8262, 2011.
\newblock \doi{10.1073/pnas.1017017108}.

\bibitem[Zenil et~al.(2018)Zenil, Kiani, and Tegn{\'e}r]{Zenil2018BDM}
Hector Zenil, Narsis~A. Kiani, and Jesper Tegn{\'e}r.
\newblock A review of graph and network complexity from an algorithmic
  information perspective.
\newblock \emph{Entropy}, 20\penalty0 (8):\penalty0 551, 2018.
\newblock \doi{10.3390/e20080551}.

\bibitem[Zenil et~al.(2019)Zenil, Kiani, Marabita, Deng, Elias, Schmidt, Ball,
  and Tegn{\'e}r]{Zenil2019}
Hector Zenil, Narsis~A. Kiani, Francesco Marabita, Yue Deng, Szabolcs Elias,
  Angelika Schmidt, Gordon Ball, and Jesper Tegn{\'e}r.
\newblock An algorithmic information calculus for causal discovery and
  reprogramming systems.
\newblock \emph{iScience}, 19:\penalty0 1160--1172, 2019.
\newblock \doi{10.1016/j.isci.2019.07.043}.

\end{thebibliography}
